\documentclass[fleqn,10pt]{wlscirep}

\usepackage{amsmath}
\usepackage{txfonts}
\usepackage{hyperref}
\usepackage{graphicx}			
\usepackage{microtype}    
\usepackage{color}
\usepackage{xcolor}
\usepackage{float}

\usepackage{lineno}
\usepackage{subcaption} 

\graphicspath{{figures/},.}

\newcommand{\ie}{i.e.,}
\newcommand{\eg}{e.g.,}

\newcommand{\expect}{\mathop{\mathrm{E}}}
\newcommand{\vect}[1]{#1}
\def\mean#1{\left< #1 \right>}

\title{Controlling congestion on complex networks: fairness, efficiency and network structure}
\author[1,*,+]{Lubos Buzna}
\author[2,+]{Rui Carvalho}

\affil[1]{University of Zilina, Univerzitna 8215/1, 01026 Zilina, Slovakia}
\affil[2]{School of Engineering and Computing Sciences, Durham University, Lower Mountjoy, South Road, Durham, DH1 3LE, UK}

\affil[*]{lubos.buzna@fri.uniza.sk}
\affil[+]{these authors contributed equally to this work}

\begin{abstract}
We consider two elementary (max-flow and uniform-flow) and two realistic (max-min fairness and proportional fairness) congestion control schemes, and analyse how the algorithms and network structure affect throughput, the fairness of flow allocation, and the location of bottleneck edges. The more realistic proportional fairness and max-min fairness algorithms have similar throughput, but path flow allocations are more unequal in scale-free than in random regular networks. Scale-free networks have lower throughput than their random regular counterparts in the uniform-flow algorithm, which is favoured in the complex networks literature. We show, however, that this relation is reversed on all other congestion control algorithms for a region of the parameter space given by the degree exponent $\gamma$ and average degree $\mean{k}$. Moreover, the uniform-flow algorithm severely underestimates the network throughput of congested networks, and a rich phenomenology of path flow allocations is only present in the more realistic $\alpha$-fair family of algorithms. Finally, we show that the number of paths passing through an edge characterises the location of a wide range of bottleneck edges in these algorithms. Such identification of bottlenecks could provide a bridge between the two fields of complex networks and congestion control.
\end{abstract}

\maketitle

\begin{document}
\section*{Introduction}
\label{sec:intro}
Twenty-first century life depends on the reliability of critical infrastructure networks. Because of high construction costs, these networks often end up operating close to the unstable region: a small increase in flow leads to network congestion or even shut-down~\cite{Ottino04,Scala14,Casals11,Carvalho15}. Without properly designed congestion control, the consequences can be catastrophic, as in the congestion collapse on the Internet~\cite{Kelly14}. In 1986, the Internet (then ARPANet) was a slow (56 Kbps) and small network with a large number of hosts (5,089)~\cite{Bidgoli04}. In October that year, the link between University of California, Berkeley and Lawrence Berkeley National Laboratory (360 m long) suffered a drop in flow rate by three orders of magnitude from 32Kbps to 40 bps. The reason for the collapse is the control mechanism implemented at the time, which focused on congestion at the receiver. The bottleneck, however, was congestion on the network. Two years later, Van Jacobson redesigned the TCP congestion control algorithm~\cite{Jacobson88}, enabling the Internet to expand in size and speed. Today, we need algorithms to share scarce network resources during times of crises~\cite{Carvalho14,Carvalho09}. In the future,  we will require algorithms to share the capacity of electrical distribution networks for the charging of electric vehicles~\cite{Carvalho15}. Moreover, when transport becomes autonomous, we may need algorithms to ease traffic congestion~\cite{Giridhar06,Helbing16}, and an understanding of the role of fairness, efficiency and network structure on such algorithms could improve the way society manages transportation. These challenges revive the topic of congestion control and uncover a new range of problems of network design at the interface between physics, engineering and the social sciences~\cite{Yu2011,Devine14,Nepusz12,Szolnoki12,Carvalho14,Carvalho15}. 
Although much work has been done to characterise congestion control mechanisms~\cite{Bertsekas92,Pioro04,Srikant04,Kelly14} and the topology of large random networks~\cite{Albert02,Boccaletti06,CaldarelliBook07,HavlinBook10,NewmanBook10}, little is known about the effect of network structure on congestion control. Furthermore, while congestion control methods have been in operation in communication networks since the 1980s, the relative performance of these algorithms on large random networks remains elusive.

\newpage

A network is at the onset of congestion when at least one edge is carrying traffic at its capacity~\cite{Kobayashi11}. When there is an attempt to increase traffic on that edge beyond its capacity, the network becomes congested, and the flow on the edge does not increase any further, even if the traffic load presented to the edge increases. 
In modelling congested complex  networks, researchers typically look for the value of a control parameter for which the network reaches the onset of congestion. Studies have focused on the onset of congestion as a function of network structure and parameters~\cite{Lai05},
optimal topologies for local search with congestion~\cite{Guimera02,Guimera02a,Cholvi05}, scaling of fluctuations in a model of an $M/M/1$ queueing system~\cite{Arenas06}, improved routeing protocols~\cite{Stanley07}, the impact of community structure on the transport of information~\cite{Danon08}, an edge weighting rule to lower costs with node capacity and increase the packet generation rate at the onset of congestion~\cite{GuanrongChen09}, and the emergence of extreme events in interdependent networks~\cite{Chen15}. These studies have the limitation that the sending frequency of packets (or rate) is uniform on the network and, consequently, the transition from free flow to congestion is determined by the nodes with the largest betweenness centrality. 
Hence, only the node(s) with the largest betweenness are fully utilised at the onset of congestion, and thus this method considerably underestimates the flow that congested networks can transport (see Methods, Section `Uniform-flow'). 
While traditionally network flows are modelled by maximising the network throughput (max-flow) or minimising the costs (minimum-cost), such efficient allocations can leave some users with zero flow, an unfair solution from the user point of view. Congestion control algorithms solve these problems by achieving cost-effective and scalable network protocols that well utilise the network capacity, sharing it among users in a fair way. These algorithms allocate path flows to paths connecting source to sink nodes. In doing so, they capture fairness by a family of  user utility functions, called $\alpha$-\textit{fair}~\cite{Mo00,Srikant04}:
\begin{equation}
U_j(f_j,\alpha)=\left\{
\begin{array}[c]{ll}
\frac{f_j^{1-\alpha}}{1-\alpha} & \text{for }\alpha\geq 0,\ \alpha\neq 1\\
\log(f_j) & \text{for }\alpha=1,
\end{array}
\right.
\label{eq:alpha_fairness}
\end{equation}
where $j=1,\ldots, R$ is a path (or user), and $f_j$ is the path flow assigned to path $j$. 
The algorithms maximise the aggregate utility $U(\alpha) = \sum_{j=1}^R U_j(f_j,\alpha)$, under the constraint that the path flows are feasible, \ie~all path flows are non-negative and no edge flow exceeds edge capacity.

For $\alpha=0$, we recover the \textit{max-flow} (MF) allocation that maximises the network throughput~\cite{Ahuja93} $U(0) = \sum_{j = 1}^{R}f_j$.   
For $\alpha=1$, we find the \textit{proportional fairness} (PF) allocation, an algorithm that manages congestion via Lagrange multipliers, which can be interpreted as an edge price. The proportional fairness optimisation problem is convex, and  Slater's qualification constraint implies that its primal and dual formulations are equivalent~\cite{Boyd04}. The primal problem is solved for the path flows, whereas the dual is solved for the Lagrange multipliers or shadow prices. Both the primal and the dual problems can be posed as decentralised optimisation problems and solved as a system of coupled ODEs~\cite{Kelly98}, which is much more efficient in large real-world networks than centralised control. Algorithmically, in the primal problem, source nodes ramp up the path flow additively but decrease it multiplicatively if at least one edge of the path is used close to capacity. The size of the system of coupled ODEs in the primal is determined by the number of paths in the network; in contrast, the number of ODEs in the dual is given by the number of network edges and is thus only dependent on network structure. Hence, if the number of paths is much larger than the number of network edges it is preferable to solve the dual instead of the primal~\cite{Kelly98,Kelly14,Carvalho14}.
The \textit{max-min fairness}  (MMF) allocation is defined by $\alpha\to \infty$ in Eq.~(\ref{eq:alpha_fairness}); it 
 is typically found, however, with a more efficient algorithm that maximises the use of network resources by users with the minimum allocation. Once these 'poor` users get the largest possible allocation, the process repeats iteratively for the next less well-off users~\cite{Bertsekas92,Carvalho12}.  Intuitively, a set of path flows is max-min fair if the wealthy can only get wealthier by making the poor even poorer.
The \textit{uniform-flow} (UF) problem is determined by the maximisation of the aggregate utility $U(\alpha)$ any $\alpha\ge 0$, with the added constraint that all path flows are the same, which implies the optimum is independent of $\alpha$. 
The  $\alpha$-fairness utility function provides a social planner with a way to understand the trade-off between efficiency ($\alpha=0$) and a continuum of models of fairness, such as proportional fairness ($\alpha=1$) and max-min fairness ($\alpha\to \infty$)~\cite{Chiu_and_Jain89,Bertsekas92,Johnson15,Kelly98,Doyle02,Massoulie02,Srikant04,Pioro04,Carvalho12}. 
The proportional fairness allocation is especial because the system and the users simultaneously maximise their utility functions, and because it is implemented in communication networks~\cite{Kelly14} (see Methods, Section `The mathematics of congestion control').

\section*{Results}
To gain insights into the behaviour of the $\alpha$-fairness family of algorithms, and to illustrate the phenomenon of congestion collapse, we first analyse the network throughput on a ring lattice. We consider a simple protocol that distributes edge capacity proportionally to flows on the paths that pass through an edge (see Methods, Section `Avoiding congestion collapse on the ring lattice'). A long path, which uses all network edges, competes for flow with a set of short paths that use only two edges each.
Individual paths may increase the flow they inject into the network with the aim of raising their edge capacity quota; queues then build up at the nodes, and the lattice becomes congested. Surprisingly, as the injected flow grows, the network throughput does not converge to an upper bound as intuitively expected, but to zero. This collapse, however, can be avoided if we control congestion with the $\alpha$-fair family of algorithms of Eq.~(\ref{eq:alpha_fairness}) . Intuitively, network throughput should decrease with an increase in $\alpha$, so that it is larger or equal for max-flow than for proportional fairness, greater or equal for proportional fairness than for max-min fairness, and in turn larger or the same for max-min fairness than for uniform-flow. In other words: we expect that the price to pay for increasing equity is a decrease in throughput, such that the proportional fairness allocation is a trade-off between efficiency (max-flow) and fairness (max-min fairness, and uniform-flow). 
Our intuition is right for small ring lattices, but as the number of nodes in the ring grows throughput in the proportionally fair and max-flow allocations converge. Indeed, proportional fairness penalises long paths because these use more network resources than short paths. As the size of the ring grows, the long path uses a higher proportion of network capacity, thus getting less and less flow, and proportional fairness converges to max-flow. In contrast, max-min fairness yields a lower throughput than these two protocols because it assigns the same allocation to all paths (see Methods, Section 'Avoiding congestion collapse on the ring lattice'). 
Hence, the ring lattice illustrates the counter-intuitive phenomena of congestion collapse, as well as, in the presence of congestion control, the surprising converge of proportional fairness to max-flow as the ring size grows. These observations made on a regular network structure with a regular structure of paths are in sharp contrast with our findings on random networks. 

\begin{figure}
\centering
\includegraphics[width=\textwidth]{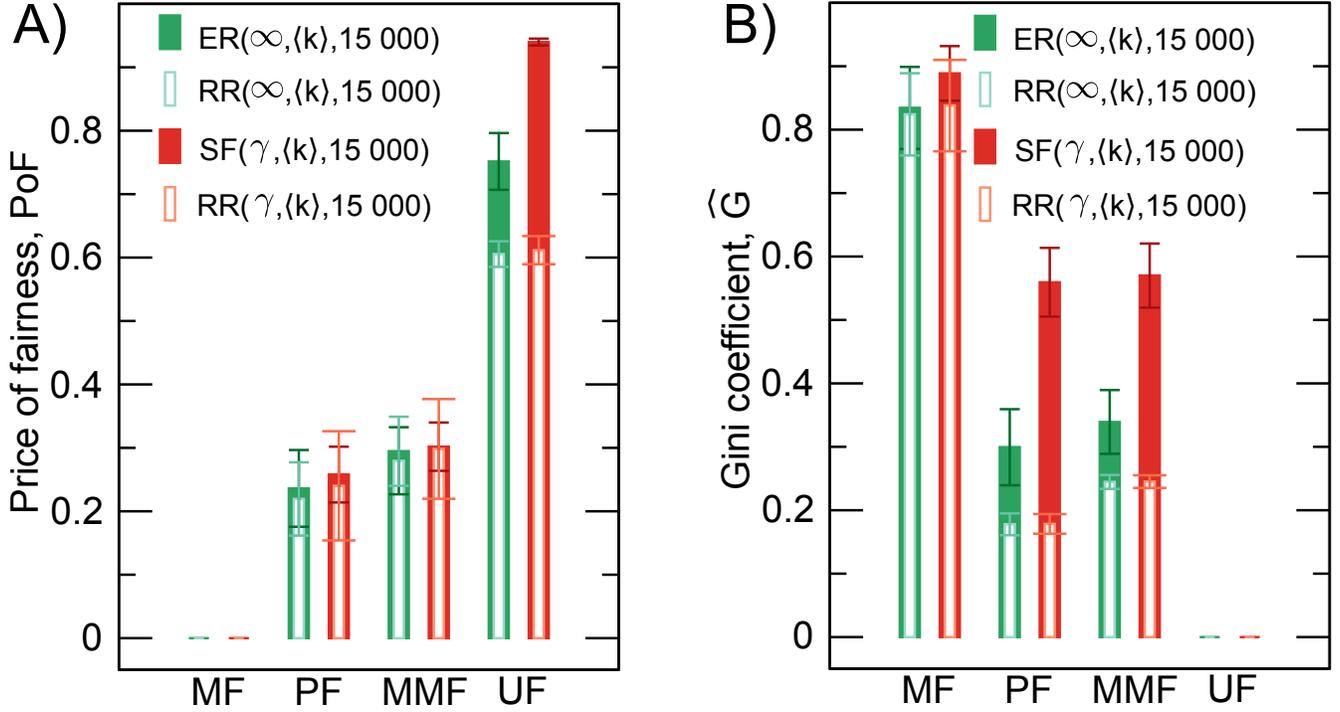}
\caption{(A) Mean of the price of fairness (PoF) and (B) Gini coefficient of path flows for each congestion control algorithm in Erd\"os-R\'enyi and scale-free networks (and their random regular counterparts) with $R=15\ 000$ paths. For each network structure, we average over 20 randomly generated networks with $N=2000$ nodes. Whiskers show the $95\%$ confidence intervals. We average the price of fairness and Gini coefficient over the mean degree $\mean{k} = \{3, 4, 5, 6, 7, 8 \}$ and the power-law exponent $\gamma = \{2.1, 2.3, 2.5, 2.7, 2.9 \}$ for scale-free $\big (\text{SF}(\gamma,\mean{k},R)\big )$ and the corresponding random regular $\big (\text{RR}(\gamma,\mean{k},R)\big )$ networks, but only over $\mean{k}$ for Erd\"os-R\'enyi $\big (\text{SF}(\gamma=\infty,\mean{k},R)\big )$ and the corresponding random regular $\big (\text{RR}(\infty,\mean{k},R)\big )$ networks.}
\label{fig:fig1}
\end{figure}

\begin{figure}
\centering
\includegraphics[width=\textwidth]{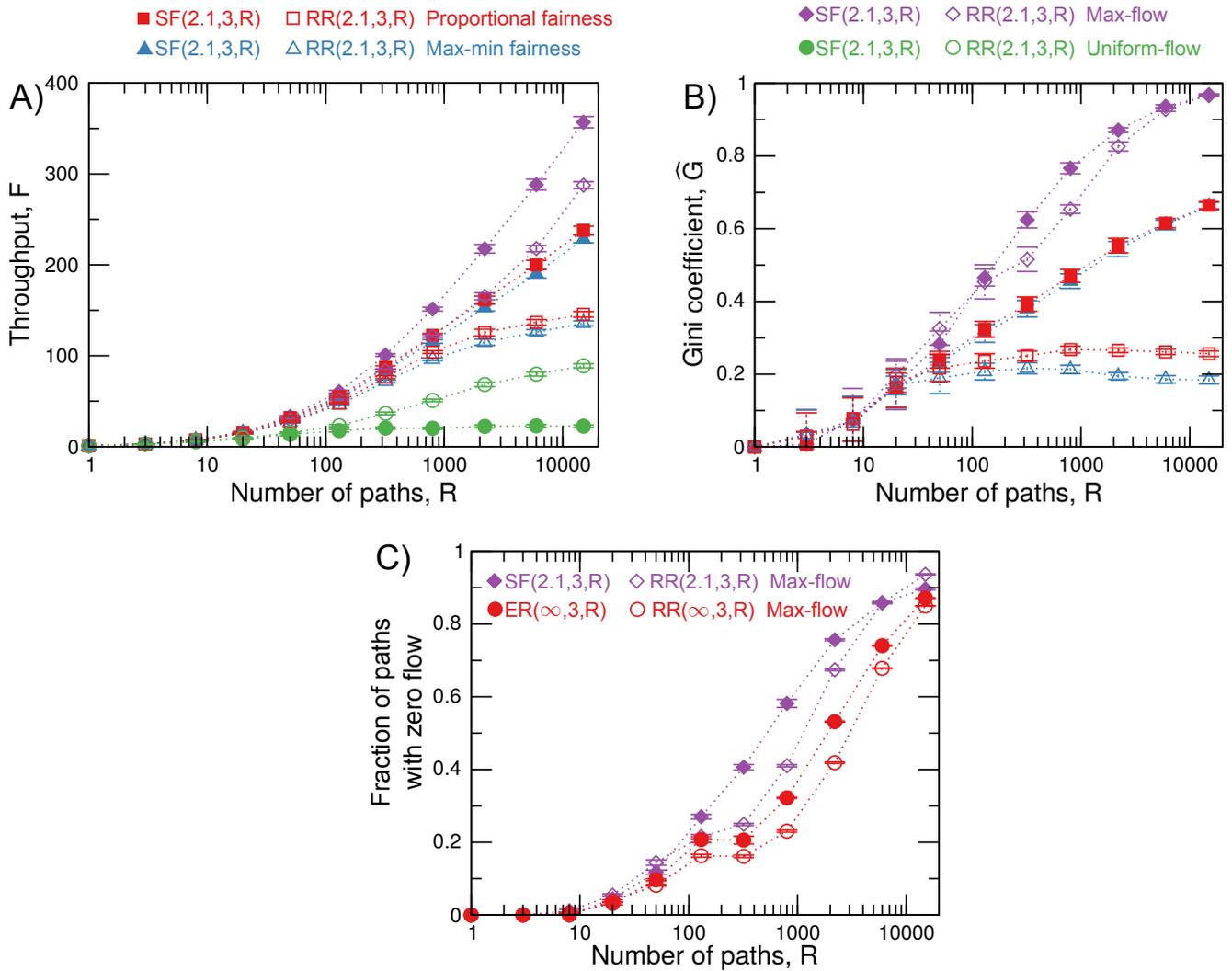}
\caption{(A) Average network throughput and (B) Average Gini coefficient of path flows in scale-free and random regular networks for $\gamma=2.1$ and $\langle k \rangle=3$ as a function of the number $R$ of shortest paths for max-flow, proportional fairness, max-min fairness,  and uniform-flow.  (C) The average fraction of paths that are allocated zero flow by the max-flow algorithm as a function of the number $R$ of shortest paths in scale-free networks $(\gamma=2.1,\langle k \rangle=3)$, Erd\"os-R\'enyi networks $(\langle k \rangle=3)$, and their random counterparts. For each network structure, we average over 20 randomly generated networks with $N=2000$ nodes. Whiskers show the $95\%$ confidence intervals.}
\label{fig:fig2}
\end{figure}

We next study the effect of controlling congestion on scale-free (SF)  (with exponent $2<\gamma<3$), Erd\"os-R\'enyi (ER), and random regular (RR) substrate networks with average node degree $3\leq k\leq 8$. Flows take place on a \textit{transport overlay network}, which is the subgraph formed by  a set of $R$ shortest paths, chosen with uniform probability among all possible shortest paths on the substrate network (see Methods, Section `Network Models'). From now on, we consider only the `transport overlay network' when we refer to random networks and omit this term from the text.
We now ask the question: to what extent is being fair compatible with maximising network throughput on random networks?

To analyse the interplay between algorithms and network structure, we next compute `the price of fairness'~\cite{Bertsimas10}, that is the relative system efficiency loss under a `fair' allocation compared to the one that maximises the sum of user utilities:
\begin{equation}
PoF(\mathcal{F},\mathcal{N}) = 1-\frac{F\left (\mathcal{F},\mathcal{N}\right )}{F\left ( \text{MF} ,\mathcal{N}\right )},
\label{eq:pof}
\end{equation}
where $\mathcal{F}\in \{\text{MF}, \text{PF}, \text{MMF}, \text{UF}\}$ is the algorithm (max-flow, proportional fairness, max-min fairness, or uniform-flow), and
$F\left (\mathcal{F},\mathcal{N}\right )$ is the throughput of the algorithm $\mathcal{F}$ for the chosen network structure. We denote the network structure by $\mathcal{N}$, such that for scale-free networks we write $\mathcal{N}:=\text{SF}(\gamma,\mean{k},R)$, and we characterise Erd\"os-R\'enyi networks ($\gamma=\infty$) by $\mathcal{N}:=\text{ER}(\infty,\mean{k},R)$. Moreover, we write $\mathcal{N}:=\text{RR}(\gamma,\mean{k},R)$ to denote the corresponding random regular networks both for scale-free and Erd\"os-R\'enyi  networks (see Methods, Section `Network Models').
The efficient algorithm (max-flow) has a price of fairness of zero, whereas an algorithm that results in zero network throughput has a price of fairness of one.
Figure~\ref{fig:fig1}A) shows that in contrast to the ring lattice, the price of fairness of proportional fairness in random networks is larger than zero, and of comparable magnitude to the price of fairness of max-min fairness, for all network structures we analysed, showing that the throughput of proportional fairness now approaches max-min fairness. To characterise the fairness of each algorithm, we show the inequality of path flows in Fig.~\ref{fig:fig1}B by the Gini coefficient (see Methods, Section `Gini coefficient'). 

An ideal congestion control algorithm would have high throughput (low price of fairness) and low inequality (low Gini coefficient) for any network structure. However, such general algorithm does not exist, because the maximisation of throughput leads to inequality. Indeed, to maximise throughput in a network with constant edge capacity, a few paths receive all the network capacity of the edges they pass through, whereas a majority of paths will be allocated zero path flow (see Fig.~\ref{fig:fig2}C). The coexistence of both types of paths leads to the vast inequality in path flows. The $\alpha$-fairness family of algorithms increases the equity of path flows with increasing $\alpha$. As a consequence, however, $\alpha$-fairness lowers network throughput as $\alpha$ increases from $\alpha=0$ (max-flow) to $\alpha=\infty$ (max-min fairness), and this mechanism captures the efficiency-fairness trade-off. 

Figures~\ref{fig:fig1}A) and B) show that the efficient algorithm (max-flow) has a high level of inequality (high Gini coefficient) and uniform-flow has low throughput (high price of fairness), thus illustrating why there is little incentive to implement these algorithms in the real-world, despite the recent interest on uniform-flow~\cite{Guimera02,Guimera02a,Lai05,Cholvi05,Arenas06,Stanley07,Danon08,GuanrongChen09,Chen15}. In contrast, proportional fairness and max-min fairness are trade-offs between efficiency and fairness, as illustrated by the mid-range values of the price of fairness and Gini coefficient for all network structures analysed. Taken together, these features uncover the effect in network throughput and fairness of elementary (max-flow and uniform-flow) versus elaborate (proportional fairness and max-min fairness) congestion control algorithms.

We observe in Fig.~\ref{fig:fig1}A that for max-flow, max-min fairness and proportional fairness, the price of fairness is largely independent of network structure. Similarly, Fig.~\ref{fig:fig1}B shows that for max-flow,  the inequality of path flows (measured by the Gini coefficient) is also largely independent of network structure. These observations suggest that proportional fairness and max-min fairness are similar algorithms with only minor dependence on network structure.  Surprisingly, however,  the inequality of path flow allocations  for  proportional fairness and max-min fairness depends mainly on network structure (see Fig.~\ref{fig:fig1}B). Hence, network designers that implement congestion control should be aware that scale-free and random regular network structures have similar throughput, but scale-free topologies induce larger inequality in path flows.
This is especially important, because proportional fairness is often implemented in real-world networks (\eg~the Internet), and the effect of network structure on the inequality of path flows is revealed by our study of the $\alpha$-fairness family of algorithms, but  
cannot be disentangled from an analysis of max-flow or uniform-flow only.
Thus, previous studies of max-flow (large inequality)\cite{Carmi2007,Carmi2008} and uniform-flow~\cite{Guimera02,Guimera02a,Lai05,Cholvi05,Arenas06,Stanley07,Danon08,GuanrongChen09,Chen15} (no inequality) miss the effect of network structure on the inequality of path flow allocations, and our study  is a natural extension to congestion control algorithms of the body of work in the complex networks literature.

To study the effect of demand on the throughput and inequality of path flows, we analyse how these quantities vary with the number $R$ of shortest paths in the network, and we study networks with $k=3$ and $\gamma=2.1$. Figure~\ref{fig:fig2}A is a plot of network throughput as the number $R$ of shortest paths grows. The Gini coefficient, plot in  Fig.~\ref{fig:fig2}B, quantifies the growth in the inequality of path flows as a function of $R$ (see Methods, Section `Gini coefficient'). 
Network throughput increases with the number of paths, since the capacity of more edges is used. Because the network size is fixed, however, the growth in throughput slows down inevitably as more paths are added to the network. The asymptotic value of throughput, and the way this slowing down takes place characterises the efficiency of the algorithm and network structure. Figure~\ref{fig:fig2}A, shows that the increase in throughput with $R$ is much slower  for uniform-flow than for the other algorithms. This result illustrates the poor performance of uniform-flow for a broad range of $R$ values and thus complements Fig.~\ref{fig:fig1}, which compares algorithms only for $R=15\ 000$.
The throughput and Gini coefficient curves  do  not intersect in Fig.~\ref{fig:fig2}, and thus the relative performance of algorithms does not change much with $R$. 

In max-flow, the path flow allocations share edge capacity on min-cuts among a relatively small number of paths, leaving most paths with zero flow (see~Fig.~\ref{fig:fig2}C), thus creating a large inequality in the assignment of path flows. Although max-flow is an extreme case, because it is the only analysed algorithm that can leave paths with zero flow, inequality is present in all congestion control algorithms. Indeed, the increase in throughput with $R$ is also accompanied by a raise in the inequality of path flow allocations in max-flow, proportional fairness and max-min fairness. 

\begin{figure}
\centering
\includegraphics[width=0.9\textwidth]{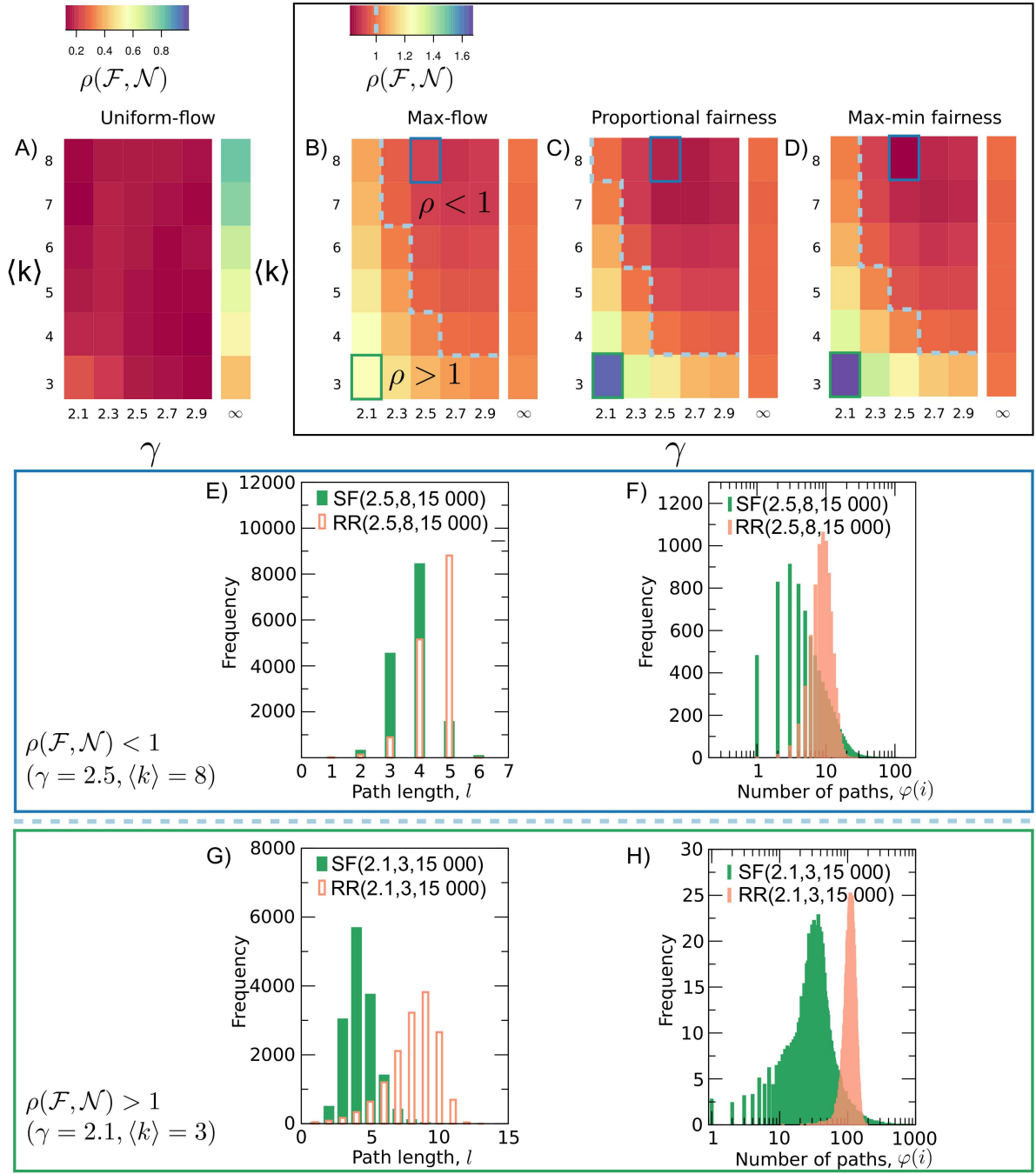}
\caption{Heatmaps of the relative throughput $\rho(\mathcal{F},\mathcal{N}) $ for uniform-flow, panel A), and the $\alpha$-fairness family of algorithms (max-flow, proportional fairness and max-min fairness), panels B) to D). We highlight the (light blue dashed) line that separates the regions $\rho(\mathcal{F},\mathcal{N} )<1$ and $\rho(\mathcal{F},\mathcal{N} )>1$. Moreover, we highlight two heatmap cells in all $\alpha$-fairness algorithms, to represent the regions $\rho(\mathcal{F},\mathcal{N}) $ smaller ($\gamma=2.5,\mean{k}=8$) and larger ($\gamma=2.1,\mean{k}=3$) than one. Panels E) to H) show the histogram of path length and the number $\varphi(\mathcal{N},i)$ of paths passing through edge $i$. Panels E) and F) examine the heatmap cell $(\gamma=2.5,\mean{k}=8)$, for which $\rho(\mathcal{F},\mathcal{N} ) <1$, whereas panels G) and H) analyse the heatmap cell $(\gamma=2.1,\mean{k}=3)$, for which $\rho(\mathcal{F},\mathcal{N} ) >1$. For each network structure, we average over 20 randomly generated networks with $N=2000$ nodes and $R=15\ 000$ shortest paths.}
\label{fig:fig3}
\end{figure}

Traditionally, congestion control algorithms have been designed to counter-balance the phenomenon that max-flow may exclude some paths (\ie~users) from using the network. However, little is known about the behaviour of these algorithms as a function of network structure. Here we take a step towards filling this gap by analysing the effect on network throughput of varying $\gamma$ and $\mean{k}$. To do this, we consider the relative throughput,
\begin{equation}
\rho(\mathcal{F},\mathcal{N} ) = \frac{F\left(\mathcal{F}, \text{SF}(\gamma,\mean{k},R)\right)}{F\left (\mathcal{F}, \text{RR}(\gamma,\mean{k},R) \right)},
\label{eq:relative_throughput}
\end{equation}
where $F(.)$ is the network throughput, $\mathcal{F}$ is the fairness algorithm, and $\mathcal{N}$ identifies the network structure and parameters ($\gamma=\infty$ denotes Erd\"os-R\'enyi and their random regular networks).
The ratio $\rho(\mathcal{F},\mathcal{N}) $ isolates the effect of node degree distribution in throughput by comparing scale-free and Erd\"os-R\'enyi networks against the null model of random regular networks (see Methods, Section `Network Models'). We observe that $\lim_{k\to (N-1)} \rho(\mathcal{F},\mathcal{N}) =1$ because both networks in the ratio converge to fully connected graphs in this limit. Together with the relative network throughput, we consider the number $\varphi(\mathcal{N},i)$ of paths passing through edge $i$:
\begin{equation}
\varphi(\mathcal{N},i) = \sum_{j=1}^R H_{ij},
\label{eq:number_paths_through_an_edge}
\end{equation}
where $H$ is the edge-path incidence matrix (see Methods, Section 'The mathematics of congestion control').  Because edge capacity is one ($c_i=1$), the path flows assigned by uniform-flow are given by $1/\max\{\varphi(\mathcal{N},i)\vert i=1,\dots,E\}$. Thus we have the exact relation between $\rho$ and $\varphi$ for uniform-flow:
\begin{equation}
\rho(\text{UF},\mathcal{N} ) =\frac{\underset{i}{\max}{\ \varphi(\text{RR}(\gamma,\mean{k},R),i)}}{\underset{i}{\max}{\ \varphi(\text{SF}(\gamma,\mean{k},R),i)}}.
\label{eq:rho_UF}
\end{equation}
We found $\rho(\text{UF},\mathcal{N}) <1$ for all $\gamma$ and $\mean{k}$, due to the higher maximum concentration of paths in scale-free networks than in random regular networks, \i.e.~$\underset{i}{\max}{\ \varphi(\text{SF}(\gamma,\mean{k},R),i)} > \underset{i}{\max}{\ \varphi(\text{RR}(\gamma,\mean{k},R),i)}$, as illustrated in Fig.~\ref{fig:fig3}A.

Similarly to uniform-flow, we could expect $\rho(\mathcal{F},\mathcal{N} ) <1$ for all values of $\gamma$ and $\langle k \rangle$. Surprisingly, however, as Figs.~\ref{fig:fig3}B-D show, $\rho(\mathcal{F},\mathcal{N}) $ can be smaller or larger than one depending on the region of parameter space $(\gamma, \langle k \rangle)$. Moreover, the dividing line $\rho(\mathcal{F},\mathcal{N} ) =1$ is largely independent of the  algorithm, indicating that network structure is the primary factor behind the relative throughput $\rho(\mathcal{F},\mathcal{N} ) $ in the $\alpha$-fairness family. 
The network structure is, however, not the only parameter influencing $\rho(\mathcal{F},\mathcal{N}) $.
To show the effect of algorithms on $\rho(\mathcal{F},\mathcal{N}) $, we analyse small values of $\gamma$ and $\mean{k}$ in Fig.~\ref{fig:fig2}A and Figs.~\ref{fig:fig3}B-D. For $\gamma=2.1$, $\mean{k}=3$ and $R=15\ 000$, throughput in max-flow is 24\% higher for scale-free than for random regular networks. For proportional fairness and max-min fairness, however, this value increases to 63\% and 68\%, respectively, disentangling flow in scale-free and random regular networks in this region of parameter space, as can be observed on the highlighted cells of the heatmaps in Fig.~\ref{fig:fig3}B-D. Figure~\ref{fig:fig2}A shows that this happens because for R = 15 000 proportionally fair and max-min fair throughput saturate in random regular networks, while throughput steadily grows with $R$ in max-flow and all algorithms in scale-free networks.

Figures~\ref{fig:fig3}B-D show the dividing line between $\rho(\mathcal{F},\mathcal{N} ) $ smaller and larger than one in parameter space. This dividing line is approximately the same in all $\alpha$-fairness algorithms. Hence, we make use of structural network measures to gain insights on system behaviour on both sides of the line.  We consider two main factors that influence network throughput. First, path length affects network throughput because paths transport a constant path flow on each of their edges. Hence, paths consume capacity from each edge they pass through, and thus the longer they are the larger the number of edges that have their available capacity reduced. Second, the pattern of path intersections influences throughput because these networks have limited edge capacity ($c=1$). Indeed, if a large number of paths pass through a limited set of edges, network throughput is restricted by the pattern of path intersections, because the limited capacity of these edges is shared among this large set of paths. In contrast, path flows and network throughput are larger if routeing is such that paths broadly avoid each other. We use $\varphi(\mathcal{N},i)$ as a simple measure to characterise the pattern of path intersections. To uncover the behaviour of path length and path intersections, we select two cells in the heatmap: $(\gamma=2.1$,$\mean{k}=3)$ to represent $\rho(\mathcal{F},\mathcal{N} ) <1$ and $(\gamma=2.5$, $\mean{k}=8)$ for $\rho(\mathcal{F},\mathcal{N} ) >1$.

To shed light on the mechanisms that explain the surprising $\rho(\mathcal{F},\mathcal{N} ) >1$ region, we show in Figs.~\ref{fig:fig3}E and G the histogram of path length, and in Figs.~\ref{fig:fig3}F and H the histograms of $\varphi(\mathcal{N},i)$ for two selected representative cells. Why do random regular networks accommodate higher flow than scale-free for $\rho(\mathcal{F},\mathcal{N} ) <1$? An analysis of the cell $(\gamma=2.5,\mean{k}=8)$ shows the probability distribution of path length is similar in scale-free and random regular networks (see Fig.~\ref{fig:fig3}E). 
However, the distribution of the number $\varphi(\mathcal{N},i)$ of paths passing through an edge $i$ is heavy-tailed for scale-free, but not for random regular networks: a relatively large number of edges are crossed by many paths in scale-free than in random regular networks (see Fig.~\ref{fig:fig3}F).  This heavy-tailed distribution of $\varphi(\mathcal{N},i)$ is an indicator of edge congestion in scale-free networks. Moreover, in random regular networks, we observe that a small number of paths pass through many edges, but not, as one would expect in congested networks, that a large number of paths pass through a limited number of edges. Hence, the distribution of $\varphi(\mathcal{N},i)$ illustrates why congestion tends to be higher in scale-free than random regular networks for $\rho(\mathcal{F},\mathcal{N}) < 1$.
Why do scale-free networks accommodate higher flow than random regular for $\rho(\mathcal{F},\mathcal{N} ) > 1$? An analysis of the cell $(\gamma=2.1,\mean{k}=3)$ shows two effects. First, paths are significantly longer in random regular ($\mean{l}= 8.1$) than in scale-free networks ($\mean{l}=4.3$). Longer paths in random regular networks consume more network resources and also intersect with other paths more often than in the region $\rho(\mathcal{F},\mathcal{N} ) < 1$, and thus will be more congested than shorter paths. Second, we only observe higher values of $\varphi(\mathcal{N},i)$ in scale-free networks than in random regular networks for a small numbers of edges. This small number of congested edges is not, however, large enough to invert the ratio $\rho(\mathcal{F},\mathcal{N} )$.  Taken together, these two effects make possible that scale-free networks accommodate larger flow than random regular for low values of $\gamma$ and $\mean{k}$.

\begin{figure}
\centering
\includegraphics[width=\textwidth]{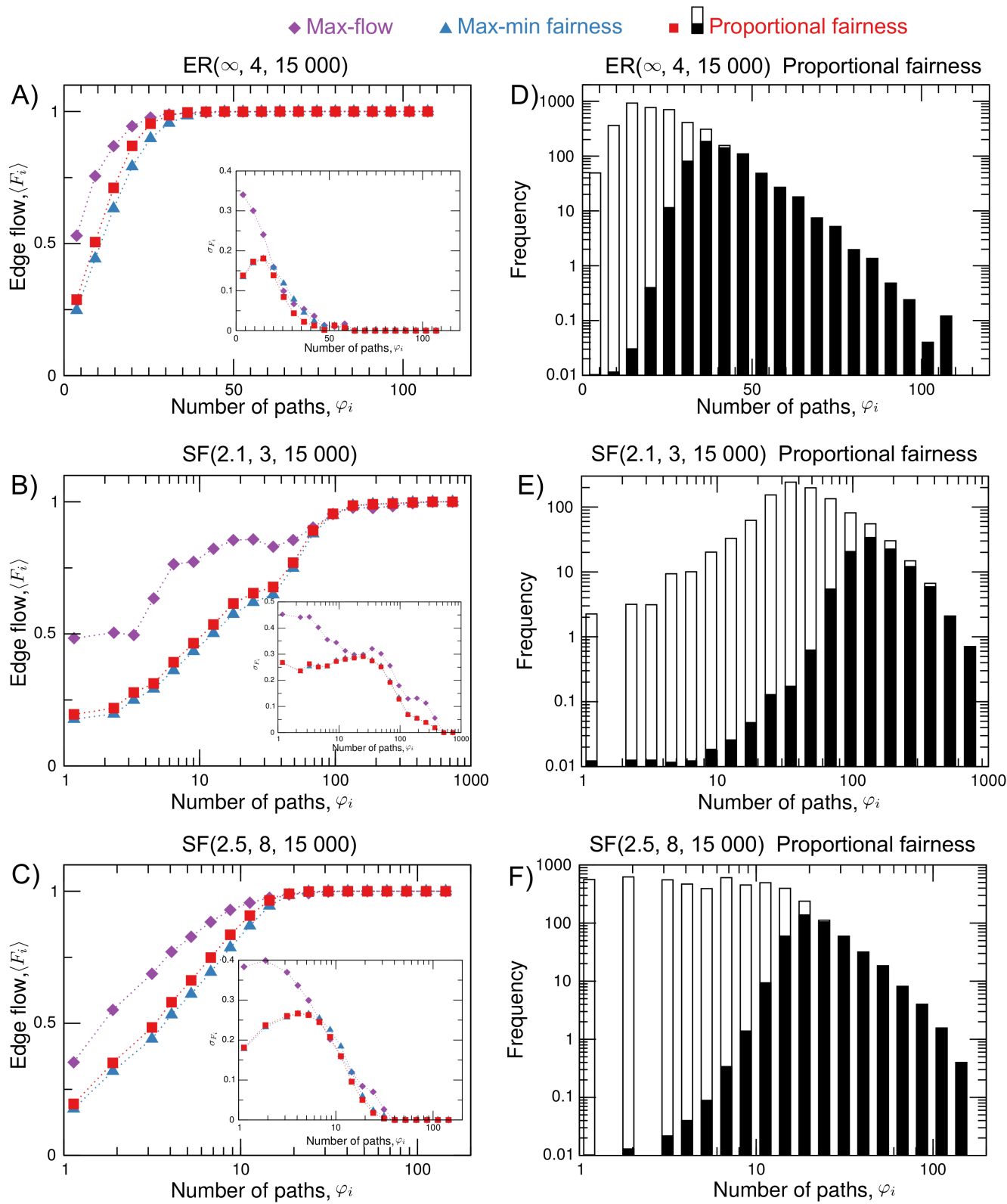}
\caption{(A-C) Average edge flow vs. the number of paths $\varphi(\mathcal{N},i)$ computed from $20$ random realisations with $N = 2000$ for Erd\"os-R\'enyi networks ER($\infty$, 4, 15 000), and for the two classes of scale-free networks on the highlighted cells in Fig.~\ref{fig:fig3}B-D, \ie~ SF(2.1, 3, 15 000) and SF(2.5, 8, 15 000). The insets show the corresponding values of the standard deviation, illustrating that the predictability of bottleneck location improves with the increase in $\varphi(\mathcal{N},i)$. Panels D-F show the histograms of $\varphi(\mathcal{N},i)$ for proportional fairness, where the shaded area of each bin is the proportion of bottleneck edges in the bin (see Fig. S1 in the Supplementary Information for max-flow and max-min fairness). Bottlenecks are saturated edges, that is edges for which $F_i \geq 0.9999c$, where $c$ is edge capacity.}
\label{fig:fig4}
\end{figure}

Currently, researchers find the onset of congestion in complex networks from betweenness centrality~\cite{Guimera02,Guimera02a,Cholvi05, Arenas06, Stanley07, Danon08, GuanrongChen09, Chen15}, a measure that captures the number of paths that cross through nodes or edges~\cite{HavlinBook10}. This uniform-flow approach finds the onset of congestion by locating the node or edge with the highest betweenness centrality, which is crossed by the largest number of paths. Figures~\ref{fig:fig1} and ~\ref{fig:fig2} illustrate, however, that uniform-flow severely underestimates throughput in the $\alpha$-fair family of algorithms at the onset of congestion, because it allocates path flows by sharing only the capacity of the most congested node or edge among the paths that pass through it. The uniform-flow algorithm thus allocates path flows globally by maximising locally the path flows that cross through the most congested edge. Hence structural measures, such as betweenness centrality, that determine the uniform-flow allocation analytically, might not be good predictors of network throughput in more realistic congestion algorithms. 

Here we are interested in the question of whether the number of paths passing through individual edges can be used to locate bottleneck edges in the $\alpha$-fair family of algorithms.
To investigate this problem, we use $\varphi(\mathcal{N},i)$ as a measure of edge load that, similarly to betweenness centrality, captures the interaction between paths on network edges.  
Edge betweenness centrality counts the number of shortest paths, which connect all possible source-sink pairs, passing through an edge. In contrast, $\varphi(\mathcal{N},i)$ counts the number of paths passing through the edge, which depends on the particular routeing used, but not just on shortest path routeing. Here, edge betweenness correlates with $\varphi(\mathcal{N},i)$, however, because we select shortest paths with uniform probability from all shortest paths.

Figures~\ref{fig:fig4}A-C show that if the number $R$ of paths is sufficiently large, edges with high value $\varphi(\mathcal{N},i)$ are used up to capacity (\ie~are bottleneck edges), with negligible standard deviation. To further relate the number of paths that cross through each edge with the location of bottleneck edges, we show in Figs.~\ref{fig:fig4}D-F the frequency of bottleneck (shaded) and non-bottleneck (clear) edges as a function of $\varphi(\mathcal{N},i)$ for proportional fairness (see Fig.~S1 of the Supplementary Information for max-min fairness and max-flow). An analysis of the shaded area on the tail of the distributions shows that a large percentage of edges with high $\varphi(\mathcal{N},i)$ are bottlenecks. For example, if we consider the 10\% of edges with the largest value of $\varphi(\mathcal{N},i)$ in SF(2.1, 3, 15 000) (ER($\infty$, 4, 15 000)), we find that on average [MF=95.3, PF=95.3, MMF=95.1]\% ([MF=99.0, PF=99.3, MMF=95.6]\%) of these are bottlenecks, representing [12.8, 21.5, 22.8]\% ([12.4, 21.5, 27.7]\%) of all bottleneck edges (the first value, enclosed in squared brackets, corresponds to max-flow, the second value to proportional fairness and the third value to max-min fairness). Thus, we find that, apart from a few mistakes, edges with high $\varphi(\mathcal{N},i)$ are bottlenecks. These results are largely independent of the congestion control algorithm (parameter $\alpha$) and of the network topology (exponent $\gamma$ and average node degree $\langle k \rangle$). 
Our numerical analysis generalises the reasoning that congestion can be characterised by the structure of paths, and can be interpreted as an extension of the analytical results for uniform-flow~\cite{Guimera02,Guimera02a,Cholvi05, Arenas06, Stanley07, Danon08, GuanrongChen09, Chen15} to the more realistic $\alpha$-fair family of congestion control algorithms combined with routeing that is determined or approximated by shortest paths.

The relation between the routeing of paths and the location of congested edges is crucial for both network designers and operators. Network designers wish to anticipate the location of bottlenecks during the design stage, so as to avoid weak links in the areas with the highest expected traffic, and to place the sensor and communication network infrastructure so as to minimise expenses with the overlaid control network. Likewise, network operators wish to determine the links that require a capacity upgrade if routeing changes. Hence, predicting the location of bottleneck edges from the routeing of paths,  may be important in real-world networks that implement congestion control algorithms (\eg~ the TCP/IP Internet congestion control protocol implements proportional fairness). 

\section*{Discussion}

We first analysed the trade-off between the efficiency and fairness of the $\alpha$-fair family of congestion control algorithms in random networks. We found that the proportional and max-min fairness algorithms generate similar throughput when results are averaged over the range of network parameters and benchmarked against the null model of random regular networks. This is significant because, in real-world systems that resemble random networks, a network operator can choose to implement proportional fairness instead of max-min fairness (the fair algorithm) with little sacrifice in fairness and throughput, and with surprisingly simple decentralised algorithms~\cite{Kelly14}. We also found that the inequality of path flows in proportional and max-min fairness depends on the structure of the network: path flows are considerably more unequal in scale-free than in random regular networks. Moreover, we showed that max-flow creates high inequality in path flow allocations and uniform-flow generates low throughput, and  thus these two algorithms are too elementary to be implemented in real-world networks.

We next characterised the growth in the network throughput and Gini coefficient as a function of the number $R$ of shortest path for a chosen scale-free network structure ($\gamma=2.1,\mean{k}=3$) and the corresponding random regular structure. 
We found that the price to pay for the increase in throughput as we independently increase $R$ or decrease $\alpha$ is an increase in the inequality of path flow allocations. We found inequality present in all algorithms, but prevalent in max-flow. Indeed, we showed that max-flow assigns zero path flow to a substantial fraction of paths, thus creating a significant inequality in the allocation of path flows. Our analysis indicates that these results are consistent across a wide range of the number $R$ of paths in the network.

Whereas this broad analysis over network parameters or a chosen network structure starts to disentangle the fairness of algorithms as a function of network structure, we next showed it is not enough to fully describe the network throughput. 
We compared the network throughput in congestion control algorithms with the null model of random regular networks in the parameter space formed by the node degree distribution exponent $\gamma$, and the average node degree $\langle k \rangle$. For the uniform-flow algorithm, we found that random regular networks, which have a more homogeneous node degree distribution than scale-free, systematically transport less flow than scale-free networks on the onset of congestion. Surprisingly, for the $\alpha$-fair family of algorithms, we found that random regular networks can support less or more flow than scale-free, depending on the region of the parameter space. Moreover, we showed that the dividing line between these two regions of parameter space can be justified by structural network measures, but that it is broadly independent of the congestion algorithm. Real-world networks could uncover further insights about the interplay between the $\alpha$ - fair family of algorithms and network topology.

An analysis of the effect of network structure based solely on uniform-flow would conclude that random regular networks have higher throughput than scale-free networks for all values of $\gamma$ and $\mean{k}$. Our results show that this conclusion is misleading. 
The uniform-flow approach leaves networks severely under-utilised in comparison with more elaborate congestion control algorithms.  We showed that uniform-flow is a crude algorithm to gain insights about the network throughput of complex networks and our findings highlight the limitations of the current line of work~\cite{Guimera02,Guimera02a,Lai05,Cholvi05,Arenas06,Stanley07,Danon08,GuanrongChen09,Chen15} on complex networks. Congestion control protocols such as max-min fairness or proportional fairness avert congestion by allocating path flows that are determined as an outcome of an optimisation procedure. Although the result is a higher level of inequality than in uniform-flow, these protocols significantly increase the network throughput and thus are superior to uniform-flow. The price to pay for elaborate algorithms for congestion control is that the rate $\lambda$ of packet production becomes source node dependent, and the critical rate is no longer found analytically. It would be hard to argue, however, that these are important factors in the modelling of real-world congested networks. Previous work on congested complex networks with uniform-flow~\cite{Guimera02,Guimera02a,Lai05,Cholvi05,Arenas06,Stanley07,Danon08,GuanrongChen09,Chen15} identifies congestion with the appearance of the first bottleneck. Inspired by this idea, we investigated whether the number of paths passing through individual edges can locate bottleneck edges in the more realistic $\alpha$-fair family of algorithms. We found that congestion on complex networks can be found not only on the edge with the largest number of paths, but on a bigger set of edges. Such edges are crossed by a high number of paths, and thus have high edge betweenness, if the routeing also follows the shortest paths.

In summary, we combined two very well established and related, but so far separated, research areas: congestion control and complex networks. We explained the main milestones in the more than 30-year old line of work in congestion control, and we compared the results from this body of literature with congestion control algorithms studied in the complex networks community in the last 15 years, which identifies the onset of congestion by considering homogeneous (uniform) path flows. On the one hand, our results show the severe limitations of the uniform-flow approach, which is the conventional algorithm to study congestion in the complex networks literature. On the other hand, we illustrated that structural characteristics typically favoured in complex networks can characterise congested edges for the $\alpha$-fair family of control algorithms, an approximation that has not received enough attention in the field of congestion control. We believe that our paper has potential to open the work in congestion control to complex networks scientists and, vice-versa that it will reveal the rich field of network science to researchers working on congestion control.
\section*{Methods}

\subsection*{The mathematics of congestion control}
Let $\mathcal{G}=(\mathcal{V},\mathcal{E})$ be an undirected and connected graph, with node-set $\mathcal{V}$ and edge-set $\mathcal{E}$, such that edge $i \in \mathcal{E}$ has capacity $c_i$. The network has $N$ nodes and $E$ edges, and a set of $R$ source and sink pairs $(s_j,t_j)$ with $s_j, t_j \in \mathcal{V}$ for $j=1,\cdots,R$. Each source and sink pair $(s_j,t_j)$ is connected by a path $r_j$, such that $\mathcal{R}=\cup_{j=1}^{R}\{r_{j}\}$ is the set of all source to sink paths on the network. The relationship between edges and paths is given by the  \textit{edge-path incidence matrix} $H$, such that $H_{ij}=1$ if edge $i$ belongs to path $r_j$, and $H_{ij}=0$ otherwise. Matrix $H$ has dimensions $E \times R$, and maps paths to the edges contained in these paths. All edges of a path $r_j$ transport the same \textit{path flow} $f_j$. The flow $F_i$ on edge $i$ is then the sum of path flows over all paths that cross the edge:
\begin{equation}
F_i = \sum_{j=1}^R H_{ij}f_j.
\label{eq_SI:total_flow_final}
\end{equation}

A vector $f$ of path flows is \textit{feasible} if  $Hf\leq c$ and $f_j \geq 0$ for $j = 1, \dots, R$, where $c$ is the vector of edge capacities. An edge is a \textit{bottleneck} if the flow passing through it is equal to the edge capacity. We define the network congestion control problem~\cite{Kelly14}: 
\begin{equation}
\begin{aligned}
& \underset{\vect f}{\text{maximise}}
&  U(\vect f )&=\sum_{j=1}^R U_j(f_j,\alpha) \\
& \text{subject to} \quad
& Hf&\leq c \\
& & f_j&\ge 0 &  \text{for } j \in 1, \dots, R,
\label{eq:flow_problem}
\end{aligned}
\end{equation}
where $\alpha \geq 0$ is a parameter and $U_j(f_j,\alpha)$ is defined by Eq.(\ref{eq:alpha_fairness}). 

In max-flow ($\alpha = 0$), to increase a path flow by $\epsilon$, we have to decrease a set of other power path flows, such that the sum of the decreases is larger or equal to $\epsilon$. In contrast, in max-min fairness ($\alpha\to \infty$), to increase a path flow by $\epsilon$, we have to decrease at least by $\epsilon$ a set of other path flows that are less or equal to the former.
Finally,  to increase a path flow by a percentage $\epsilon$ in proportional fairness ($\alpha = 1$), we have to decrease a set of other power path flows, such that the sum of the percentage decreases is larger or equal to $\epsilon$~\cite{Kelly14}.

\subsubsection*{Max-min fairness}
Formally, a vector $f$ of path flows is max-min fair, if it is feasible and if for any other feasible vector $f^{\prime}$ of path flows, there exists a path  $r_j\in \mathcal{R}:f_j^{\prime}>f_j$ implies that there exists another path $r_l\in \mathcal{R}:f_l^\prime< f_l$ and $f_l\leq f_j$~\cite{Pioro04}. The max-min fairness allocation is the solution of problem (\ref{eq:flow_problem}) for $\alpha \rightarrow \infty$. The allocation is typically found, however, with an iterative algorithm~\cite{Pioro04} that locates the bottleneck edges. The algorithm first increases all path flows uniformly from zero until it maximises the smallest path flows, that is until it finds the first bottleneck edges. The path flows on paths that pass through these bottlenecks cannot be increased because the edges are used to their full capacity, and hence the algorithm fixes these path flows, and updates the residual capacity  still available to other paths. Next, the process is repeated for the paths that do not have yet a fixed path flow. To describe the algorithm formally, we define $\mathcal{R}^{(m)}$ to be the set of paths on the network at iteration $m$, and $\mathcal{R}^{(m)}_i$ to be the subset of paths in $\mathcal{R}^{(m)}$ that cross through edge $i$. Before we start the algorithm, we assign $\mathcal{R}^{(1)}=\mathcal{R}$ and $c_i^{(1)}=c_i$ for all edges, and a path flow $f^{(0)}_j=0$ to each path $r_j\in \mathcal{R}^{(1)}$. Next, we initialise the iteration counter $m=1$. In the first step of the MMF  algorithm, for each edge $i$ with non-zero capacity that belongs to at least one path, we define the edge capacity divided equally among all paths that pass through the edge at iteration $m$ of the algorithm as:
\begin{equation}
s^{(m)}_i=c^{(m)}_i/\left\vert \mathcal{R}^{(m)}_i\right\vert,
\label{eq_SI:phi}
\end{equation}
for all $c^{(m)}_i\neq 0$. We then find the minimum of $s^{(m)}_i$, given by
\begin{equation}
\Delta f^{(m)}=\min_{i\in \mathcal{E},c^{(m)}_i\neq 0}s^{(m)}_i.
\label{eq_SI:path_flow_at_step_j}
\end{equation}
In the second step of the MMF  algorithm, we increase all path flows of paths in $\mathcal{R}^{(m)}$ by $\Delta f^{(m)}$, such that
\begin{equation}
f_j^{(m)}=\left\{
\begin{array}
[c]{ll}
f_j^{(m-1)}+\Delta f^{(m)} & \text{if }r_j\in \mathcal{R}^{(m)}\\
f_j^{(m-1)} & \text{if }r_j\in \mathcal{R}\backslash \mathcal{R}^{(m)}%
\end{array}
\right.  .
\end{equation}
The effect is to saturate the set of bottleneck edges $\mathcal{E}_B^{(m)} =\{i \in \mathcal{E}:\sum_{j=1}^R H_{ij} \Delta f^{(m)}=c^{(m)}_i\}$, and consequently also to saturate the set of paths that contain at least one bottleneck edge. Next, we create a residual network, by subtracting the capacity used by the path flows,
\begin{equation}
c^{(m+1)}_i=c^{(m)}_i-\sum_{j=1}^R H_{ij} \Delta f^{(m)}.
\label{eq_SI:capacity_update}
\end{equation}
Note that all bottleneck edges will be saturated, that is each will have $c^{(m+1)}_i=0$ after this step. We also say that all paths that contain at least one bottleneck edge are saturated paths, to mean that their path flow will not be increased in subsequent iterations of the MMF algorithm. We say that $\mathcal{R}^{(m+1)}$ is the set of augmenting paths because the path flows of paths in $\mathcal{R}^{(m+1)}$ can still be increased in subsequent iterations of the algorithm, and update it following:
\begin{equation}
\mathcal{R}^{(m+1)}   = \mathcal{R}^{(m)}\backslash \cup_{i\in \mathcal{E}_{B}^{(m)}}\mathcal{R}^{(m)}_{i}.
\label{eq_SI:paths_update}
\end{equation}
Finally, if $\mathcal{R}^{(m+1)}$ is not empty, we increase the iteration counter $m\leftarrow m+1$, and go back to the first step, otherwise we stop. 

\subsubsection*{Proportional fairness}
A vector of path flows ${\vect f^\ast}=(f_1^\ast,\ldots,f_R^\ast)$ is \textit{proportionally fair} if it is feasible and if for any other feasible vector of path flows ${\vect f}$, the sum of proportional changes in the path flows is non-positive~\cite{Kelly98,Tan99}:
\begin{equation}
\sum_{j=1}^{R}\frac{f_j-f_j^\ast}{f_j^\ast}\leqslant 0.
\label{eq_SI:proportional_fairness}
\end{equation}
The proportionally fair allocation is found from problem (\ref{eq:flow_problem}) with the utility function in Eq.~(\ref{eq:alpha_fairness}) for $\alpha=1$, and we refer to this problem as the primal~\cite{Kelly14}. The optimisation problem is convex because the aggregate utility $U(\vect f )$ is concave and the inequality constraints are convex. Thus, any locally optimal point is also a global optimum, and we can use results from the theory of convex optimisation to find the proportional fair flow allocation (see~\cite{Courant89} and ~\cite{Ball_PCM08} for a brief introduction to Lagrange multipliers, and~\cite{Boyd04} on convex optimisation).
The Lagrangian is given by~\cite{Kelly98,Tan99}:
\begin{equation}
L(\vect f, \vect \mu)=\sum_{j=1}^R \log(f_j)+\vect \mu^T(\vect c-H\vect f),
\label{eq:Lagrangian}
\end{equation}
where $\mu=(\mu_1,\ldots,\mu_E)$ is a vector of Lagrange multipliers. The Lagrange dual function~\cite{Boyd04} is then given by $\sup_{f}L(f,\mu)$, which is easily determined analytically by $\partial L(f^\ast,\mu^\ast)/\partial f=0$ as
\begin{align}
\frac{\partial L(f^\ast,\mu^\ast)}{\partial f_{j}^\ast} & =\frac{1}{f_{j}^\ast}-\sum_{i=1}%
^E H_{ij}\mu_i^\ast=0\Leftrightarrow f_{j}^\ast =\frac{1}{\sum_{i=1}^E H_{ij}\mu_i^\ast}  &  \text{for }  j \in 1, \dots, R,
\label{eq:Lagrange_dual}
\end{align}
and thus
\begin{equation}
\sup_{f}L(f,\mu)=-\sum\limits_{j=1}^R\log\left(  \sum\limits_{i=1}^E H_{ij}\mu_i\right)
+\sum\limits_{i=1}^E \mu_ic_{i}-R.
\label{eq:Lagrange_dual_final}
\end{equation}
After removing the constant term in equation~(\ref{eq:Lagrange_dual_final}) and converting to a maximisation problem, we obtain the dual problem~\cite{Kelly98,Tan99}
\begin{equation}
\begin{aligned}
& \underset{\vect \mu}{\text{maximise}}
& V(\mu) &= \sum\limits_{j=1}^R\log\left(  \sum\limits_{i=1}^E H_{ij}\mu_i\right)
-\sum\limits_{i=1}^E \mu_ic_i \\
& \text{subject to} \quad
& \mu_i&\geq 0  &  \text{for } i \in 1, \dots, E,
\label{eq:dual}
\end{aligned}
\end{equation}
where $\mu=(\mu_1,\ldots,\mu_E)$ is a vector of dual variables. The primal problem is convex and the inequality constraints are affine. Hence, Slater's condition is verified and thus strong duality holds. This means that the \textit{duality gap}, \ie~the difference between the optimal of the primal problem~(\ref{eq:flow_problem}) and the optimal of the dual problem~(\ref{eq:dual}), is zero~\cite{Boyd04}. The primal objective function depends on $R$ variables (the path flows) and is constrained by an affine system of equations, whereas the dual objective function depends on $E$ variables (the edges) and is constrained only by the condition that the dual variables are non-negative. Thus, the dual problem~(\ref{eq:dual}) is more efficient to solve than the primal when the number of paths exceeds number of network edges. The optimal path flows can then be recovered from the optimal Lagrange multipliers with Eq.~(\ref{eq:Lagrange_dual}). 

The decentralized implementation of proportional fairness relies on a feed-back mechanism on path flows~\cite{Kelly14}: multiplicatively decrease path flows of paths that pass through bottlenecks and additively increase all other path flows. The combination of the fast correction (multiplicative decrease) and slow ramp-up (additive increase) is the mechanism behind the TCP internet congestion control protocol. Crucially, this mechanism requires each bottleneck to send a feedback signal to the sender of each path, with the information that the path flow should be additively increased or multiplicatively decreased. Knowledge of where to place sensors and where to connect to the communication network that sends the feedback signals is thus important for the network designer and operator.
\subsubsection*{Uniform-flow}

The uniform-flow allocation can be found for any $\alpha\ge 0$ from problem (\ref{eq:flow_problem}) and Eq.~(\ref{eq:alpha_fairness}), with an additional set of constraints ensuring that path flows are uniform:
\begin{equation}
\begin{aligned}
& &  f_1& = f_j & \text{for } j\in 2, \dots, R.\\
\label{uniform_flow_problem}
\end{aligned}
\end{equation}
The optimal uniform-flow allocation is $\alpha$-invariant, because $U_j(f_j,\alpha)$ is a monotonically increasing function of $f_j$ for any $\alpha \geq 0$. Algorithmically, the uniform-flow allocation can also be found as the solution to the first iteration ($m=1$) of the max-min fairness algorithm, since the algorithm maximises the minimum path flow allocation, and all path flows are the same at the end of the first iteration.

The onset of congestion in complex networks is often determined by the uniform-flow allocation~\cite{Guimera02,Guimera02a,Lai05,Cholvi05,Arenas06,Stanley07,Danon08,GuanrongChen09,Chen15}. At each time step, source node $n$ generates a packet with probability $\lambda$ and sends it towards the sink node along a shortest path. The expected number of packets in the network at each time step is $\lambda N D$, where $D$ is the average shortest path length. Moreover, the probability that a packet will pass through a node $n_{max}$ with the largest betweenness is $B_{n_{max}}/\sum_{n=1}^N B_n$ (here, the betweenness centrality $B_n$ of a node $n$ equals the number of shortest paths between all pairs of nodes in the network going through node~$n$~\cite[page 28]{HavlinBook10}). The average number of packets that node $n_{max}$ receives at each time step is thus  $Q_{in}=\lambda D B_{n_{max}}/\left ( (N-1) D \right )$, where we used the simplification that the sum of the betweenness values of all nodes is the number of pairs of nodes on the network multiplied by the average path length, $\sum_{n=1}^N B_n= N(N-1) D$. At each time step, the node with the highest betweenness can deliver $Q_{out}=c_{n_{max}}$ packets, and hence the onset of congestion is given by $Q_{out}=Q_{in}$, that is 
\begin{equation}
\lambda_c =c_{n_{max}}(N-1)/B_{n_{max}}.
\label{eq:lambda}
\end{equation}
This deduction considers a network of capacitated nodes, but we can have capacity constraints on the links instead, and packets may queue at the nodes for service. Congestion control algorithms are similar for node and link capacity, and here we analyse random networks with link capacity, because this is the standard in the modelling of communication networks~\cite{Kelly14}.

The reasoning leading to Eq~(\ref{eq:lambda}) assumes that the number $\lambda_c/(N-1)$ of packets injected into the network per path at each time step is the same for all paths, and that it is determined by the ratio between the node capacity $c_{n_{max}}$ and the number of paths $B_{n_{max}}/(N-1)$ passing through node $n_{max}$. The obvious drawback of this approach is that the estimate of $\lambda_c$ in Eq.~(\ref{eq:lambda}) considers only the first bottleneck to appear in the network, and thus underestimates the load typically present in congested networks. 
 
\subsection*{Avoiding congestion collapse on the ring lattice}
\label{ring_lattice}

\begin{figure}
\centering
\includegraphics[width=0.5\linewidth]{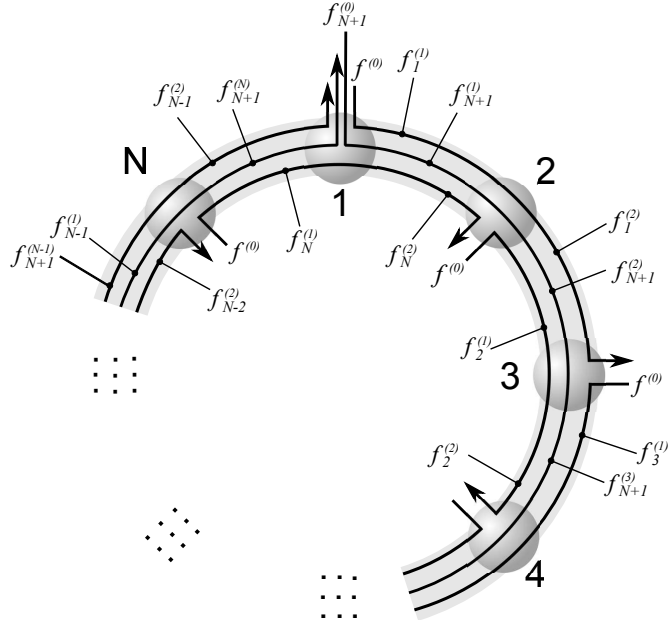}
\caption{Ring lattice with $N$ nodes and edges, such that each edge has capacity $c$, with paths and nodes indexed clockwise from the top. Short paths inject a flow $f^{(0)}$ at each node and use the network for two edges before exiting, and a long path injects a flow $f_{N+1}^{(0)}$ at node $1$ and uses all edges of the network. If we distribute edge capacity proportionally to flows entering the edge, the network throughput goes to zero as the flow injected at each node increases, because queues of increasing length build up at each node.}
\label{fig:fig5}
\end{figure}

Consider a ring lattice of $N$ nodes,  each connected to its nearest neighbours by an edge with finite capacity $c$,  as illustrated in Fig.~\ref{fig:fig5}. We relax the constraint that all edges of path $r_j$ transport the same path flow $f_j$, allow instead queues to build up at the nodes, and thus edge flows to differ on edges along a path. User $j$ injects a flow $f^{(0)}$ into the network at node $j\in \{1,\ldots,N\}$ on a \textit{short} path $j$; the flow $f_j^{(1)}$ passes from node $j$ to node $(j+1)\pmod{N}$; the flow $f_j^{(2)}$  passes from node $(j+1)\pmod{N}$   to node $(j+2)\pmod{N}$ and exits the network. Consider also that node $1$ is a source and sink of a \textit{long} path, user $N+1$, that passes through all nodes, with flow $f_{N + 1}^{(j)}$ on the edge linking node $j$ to node $(j+1)\pmod{N}$. The subscript $j$ identifies the short path, as well as its first node. The superscripts $(1)$ and $(2)$ index the edges the short paths pass through, and superscript $(j)$ indexes edges the long path cross.

To illustrate the mechanism of congestion collapse, we assume a simple congestion control scheme that distributes the total flow  $F_j=f^{(0)} + f_{(N + j -2) \pmod N + 1}^{(1)} + f_{N + 1}^{(j-1)}$ at node $j$ proportionally to the paths that pass through the node:

\begin{align} 
f_j^{(1)} &= \min \left ( f^{(0)}, \frac{f^{(0)} }{F_j}\ c \right ) \label{eq_f_1} \\
f_{(N + j -2) \pmod N + 1}^{(2)} &= \min  \left (f_{(N + j -2) \pmod N + 1}^{(1)}, \frac{f_{(N + j -2) \pmod N + 1}^{(1)}}{F_j}\ c\right) \label{eq_f_2}\\
f_{N + 1}^{(j)} &= \min \left ( f_{N+1}^{(j-1)}, \frac{f_{N+1}^{(j-1)}}{F_j}\ c\right ). \label{eq_f_3}
\end{align}
The network is not congested for $f^{(0)} <\left (c-f_{N+1}^{(0)}\right )/2$, and in this case the throughput is $Nf^{(0)}+f_{N+1}^{(0)}$. If the network is congested, flows decrease along a path, that is $f^{(0)}>f_j^{(1)}>f_j^{(2)}$, and queues build up at each node. The proportional allocation of edge capacities may motivate individual users to increase  $f^{(0)}$ in order to receive a larger share of network capacity. However, as the flow $f^{(0)}$ injected at each short path grows, the length of queues at the nodes also grow and the network throughput decreases. This \textit{congestion collapse} is a consequence of the collapse of throughput both for short and long paths. Indeed, in the limit $\{ f^{(0)}, f_{N+1}^{(0)}\} \rightarrow \infty$, the system of Eqs.~(\ref{eq_f_1}-\ref{eq_f_3}) yields $f_{1}^{(1)} = c/2$, $f_{j}^{(1)} = c$ for $j\geq 2$ and $f_{j}^{(2)} = 0$ for all short paths;
and $f_{N+1}^{(1)} = c/2$, and $f_{N + 1}^{(j)} = 0$ for $j\geq 2$ on the long path. Hence, $\sum\limits_{j = 1}^{N} f_j^{(2)} + f_{N+1}^{(N)} \rightarrow 0$. 

Let us now assume the more restrictive condition that $N\geq 2$ paths carry a path flow, and consider the effect of controlling congestion. Because 
the path flow is  constant on all edges along a path, we have a path flow $f$ for every short path, and a path flow $f_{N+1}$ for the long path, such that $f_{N+1}+2f=c$ at every edge, and thus
\begin{equation}
f=(c-f_{N+1})/2.
\label{eq:pf_modified_parking_lot}
\end{equation}
One max-flow allocation is $f=c/2$ and $f_{N+1}=0$, leaving user $N+1$ with no access to the network, with a network throughput of $Nc/2$.
The max-min fair  solution is $f=f_{N+1}=c/3$, with a network throughput of $(N+1)c/3$.
The proportionally fair solution is found by maximising $U=\log f_{N+1}+\sum_{j=1}^{N}\left ( \log (c-f_{N+1})-\log 2\right )$ over $f_{N+1}$, yielding the path flow on the long path:
\begin{equation}
f_{N+1}^\ast=\frac{c}{N+1}.
\label{eq:f_PF_long}
\end{equation}
Combining Eqs.~\ref{eq:pf_modified_parking_lot} and \ref{eq:f_PF_long} yields the proportionally fair path flow on the short paths:
\begin{equation}
f^\ast=\frac{N}{2(N+1)}c\label{eq:f_PF_short}.
\end{equation}
Hence, the proportional fair network throughput is $\left (N^2+2\right )c/\left (2\left (N+1\right )\right )$. As the size of the network increases, the proportional fair allocation approaches the max-flow solution, leaving the long path with ever smaller flow. In contrast, the max-min fair protocol assigns the same allocation to all paths independently of network size, at the expense of having a lower throughput than proportional fairness and max-flow. Thus, this example shows that proportional fairness and max-flow generate the same throughput on an infinitely size ring lattice, and that this is higher than the throughput provided by max-min fairness.

\subsection*{Network Models}

We are interested in global congestion patterns, and thus require connected networks.
We generate undirected, unweighted (\ie~unit capacity) and connected scale-free (SF) networks following the static model~\cite{Goh01}, and call these the \textit{substrate networks}. We start with $N=2000$ disconnected nodes and assign a weight $w_i=i^{-\beta}$ to each node $i$ ($i=1,\dots,N$), where $\beta\in  [0,1)$. We randomly select two nodes $i$ and $j$ with probability proportional to $w_i$ and $w_j$, respectively, and connect them if they are not yet connected, avoiding self-loops and multiedges. We repeat this procedure until the average node degree of the largest connected component is $\mean{k}$, and keep only the largest connected component. The degree distribution follows a power-law with exponent $\gamma=(1+\beta)/\beta$. We generate scale-free networks with average degree $\mean{k}\in \left \{3,4,5,6,7,8 \right \}$, and $\gamma=\left \{ 2.1,2.3,2.5,2.7,2.9,3.1\right \}$.
We treat Erd\"os-R\'enyi networks as a special case of scale-free networks for $\beta=0$ ($\gamma=\infty$). 
This procedure generates networks with different number of nodes and edges, dependent on $\gamma$ and $\mean{k}$. To overcome this limitation, we compare each network generated by the static model with a connected random regular (RR) graph with the same average degree as the scale-free (SF) or Erd\"os-R\'enyi (ER) network. This RR network has the same number of nodes and edges as the corresponding SF network generated with the static model, and thus can be seen as a rewired graph, such that each node has a fixed degree. We then use the RR network as a null model, against which we analyse the features of the corresponding SF or ER network. 

In real-world flow networks, most flow between a pair of source and sink nodes will be located over only one route~\cite{NacePioro08}, and that is typically the shortest path because it minimises the cost of transport~\cite{Danila2006,Barthelemy11}. Researchers have explored a variety of alternatives to shortest path routeing~\cite{Goh05,Danila2006,Stanley07}, yet there is no clear alternative to shortest path routeing from this effort and these algorithms have often been designed for specific models and scale-free networks. 
An alternative way to determine the routeing would be to find all elementary paths (paths that do not traverse any node more than once) between source and sink node pairs, but this is only practical for small networks because the number of paths grows exponentially with network size\cite{Piororeview14}.
Hence, we analyse routeing between a source and sink pair along shortest paths only. We chose $R$ shortest paths with uniform probability from the set of all shortest paths, and  extract the \textit{transport overlay network}~\cite{Wang08} composed of the edges that are crossed by at least one path on the substrate network. This transport overlay network is the union of all $R$ shortest paths, and is the subgraph of the substrate network that carries flow.

\subsection*{Gini coefficient}

To characterise inequalities in the flow allocations, we analyse the Gini coefficient of path flows. The Gini coefficient is defined as~\cite{Ullah98}
\begin{equation}
 G=\frac{1}{2\mu} \expect \left [\left | u - v \right | \right ]=\frac{1}{2\mu}\int_{0}^{\infty}\int_{0}^{\infty}\left\vert u-v\right\vert g(u)g(v)\ du\ dv,
\label{eq:gini_coefficient_continuous}
\end{equation}
where $u$ and $v$ are independent identically distributed random variables with probability density $g$ and mean $\mu$. In other words, the Gini coefficient is one half of the mean difference in units of the mean. The difference between the two variables receives a small weight in the tail of the distribution, where $g(u)g(v)$ is small, but a relatively large weight near the mode. Hence, $G$ is more sensitive to changes near the mode than to changes in the tails.
For a random sample ($x_l$, $l=1,2, \ldots, n$), the empirical Gini coefficient,~$\widehat{G}$, may be estimated by a sample mean
\begin{equation}
 \widehat{G}=\frac{\sum_{l=1}^n\sum_{o=1}^n \left\vert x_l - x_o \right\vert}{2n^2\mu}.
\label{eq:gini_coefficient_discrete}
\end{equation}
The Gini coefficient is used as a measure of inequality, because a sample where the only non-zero value is $x$ has $\mu=x/n$ and hence $\widehat{G}=(n-1)/n \to 1$ as~$n\to\infty$, whereas $\widehat{G}=0$ if all data points have the same value.

\section*{Acknowledgements}
We thank Matej Cebecauer for conducting numerical experiments and analysing results in early versions of the manuscript. We thank Dirk Helbing for granting access to the ETHZ Brutus high-performance cluster. This work was supported by the Alan Turing Institute, call for collaboration in the Lloyd’s Register Foundation Programme to support data-centric engineering under grant number LRF16-05, by the Engineering and Physical Sciences Research Council under grant number EP/I016023/1, by VEGA (project 1/0463/16), APVV (project APVV-15-0179) and by FP 7 (project ERAdiate 621386).

\section*{Author contributions statement}
L.B. and R.C conceived the experiment(s),  L.B. and R.C wrote the manuscript. 

\section*{Additional information}
Competing financial interests: The authors declare no competing financial interests.

\newpage

\section*{Supplementary Information: Controlling congestion on complex networks: fairness, efficiency and network structure}

\begin{center}
{\bf Lubos Buzna and Rui Carvalho}
\end{center}

\begin{figure}[H]
\centering
\includegraphics[width=0.85\textwidth]{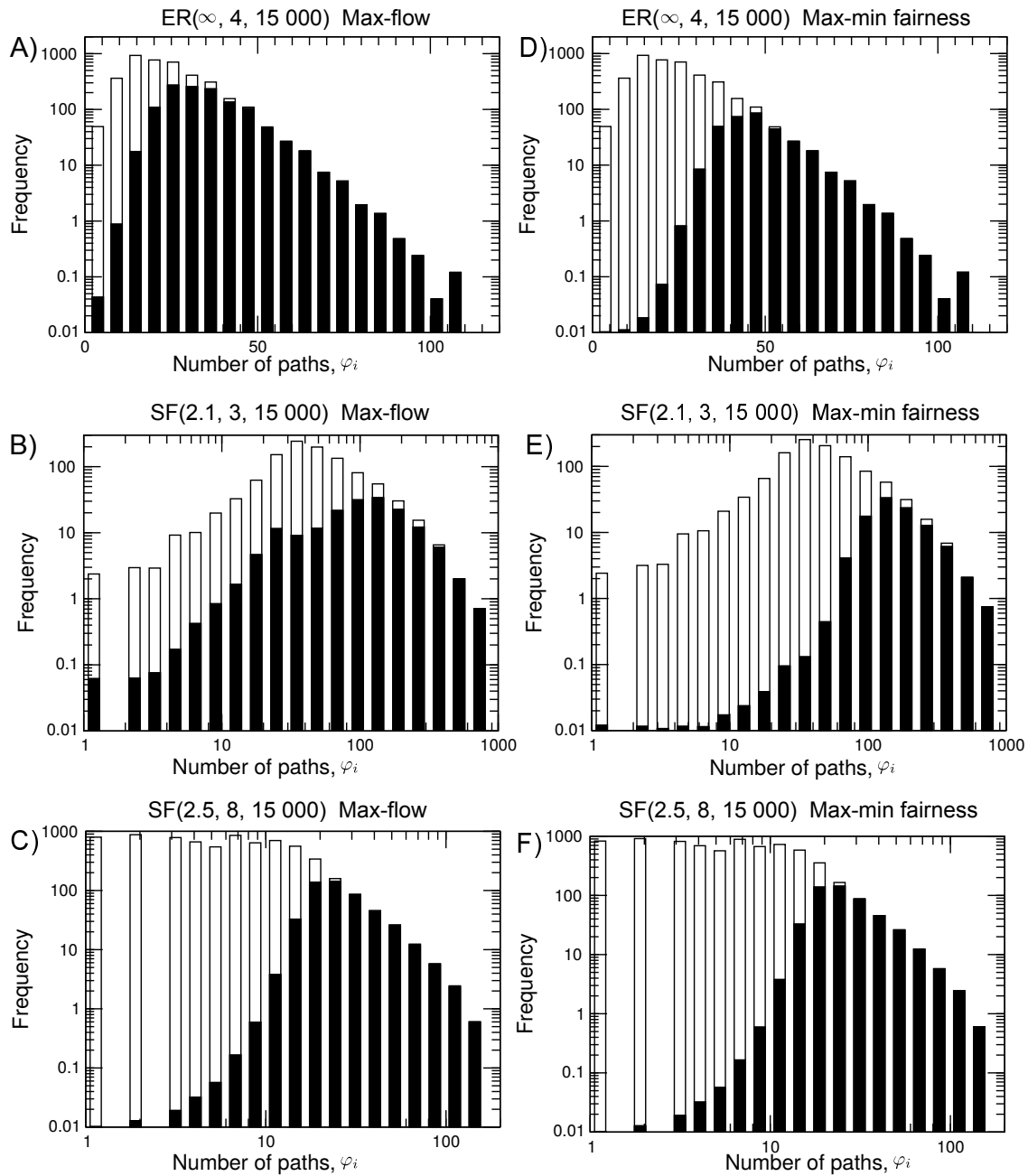}
\caption{(A-C) Histograms of  $\varphi(\mathcal{N},i)$ values computed from $20$ random realisations and $N = 2000$ for ER($\infty$, 4, 15 000) networks and two classes of SF networks corresponding to the two highlighted cells in Fig.~3B-D, \ie~SF(2.1, 3, 15 000) and SF(2.5, 8, 15 000). Panels A-C (D-E) show histograms for the max-flow (max-min fairness), where the shaded area of each bin is the proportion of bottleneck edges in the bin. Bottlenecks are saturated edges, that is edges for which $F_i \geq 0.9999c$, where $c$ is edge capacity.}
\label{fig:figSI1}
\end{figure}

\begin{thebibliography}{10}
\expandafter\ifx\csname url\endcsname\relax
  \def\url#1{\texttt{#1}}\fi
\expandafter\ifx\csname urlprefix\endcsname\relax\def\urlprefix{URL }\fi
\providecommand{\bibinfo}[2]{#2}
\providecommand{\eprint}[2][]{\url{#2}}

\bibitem{Ottino04}
\bibinfo{author}{Ottino, J.~M.}
\newblock \bibinfo{title}{{Engineering complex systems}}.
\newblock \emph{\bibinfo{journal}{Nature}} \textbf{\bibinfo{volume}{427}},
  \bibinfo{pages}{399--399} (\bibinfo{year}{2004}).

\bibitem{Scala14}
\bibinfo{author}{Scala, A.} \emph{et~al.}
\newblock \emph{\bibinfo{title}{{Power Grids, Smart Grids and Complex
  Networks}}}, \bibinfo{pages}{97--110}.
\newblock {NATO Science for Peace and Security Series C: Environmental
  Security} (\bibinfo{publisher}{Springer Netherlands}, \bibinfo{year}{2014}).

\bibitem{Casals11}
\bibinfo{author}{Rosas-Casals, M.} \& \bibinfo{author}{Sol{\'e}, R.}
\newblock \bibinfo{title}{{Analysis of major failures in Europe{\rq}s power
  grid}}.
\newblock \emph{\bibinfo{journal}{International Journal of Electrical Power \&
  Energy Systems}} \textbf{\bibinfo{volume}{33}}, \bibinfo{pages}{805--808}
  (\bibinfo{year}{2011}).

\bibitem{Carvalho15}
\bibinfo{author}{Carvalho, R.}, \bibinfo{author}{Buzna, L.},
  \bibinfo{author}{Gibbens, R.} \& \bibinfo{author}{Kelly, F.}
\newblock \bibinfo{title}{{Critical behaviour in charging of electric
  vehicles}}.
\newblock \emph{\bibinfo{journal}{New J. Phys.}} \textbf{\bibinfo{volume}{17}},
  \bibinfo{pages}{095001} (\bibinfo{year}{2015}).

\bibitem{Kelly14}
\bibinfo{author}{Kelly, F.} \& \bibinfo{author}{Yudovina, E.}
\newblock \emph{\bibinfo{title}{{Stochastic Networks}}}
  (\bibinfo{publisher}{Cambridge University Press}, \bibinfo{year}{2014}).

\bibitem{Bidgoli04}
\bibinfo{author}{Bidgoli, H.}
\newblock \emph{\bibinfo{title}{{The Internet Encyclopedia}}},
  vol.~\bibinfo{volume}{II} (\bibinfo{publisher}{John Wiley \& Sons Inc},
  \bibinfo{year}{2004}).

\bibitem{Jacobson88}
\bibinfo{author}{Jacobson, V.}
\newblock \bibinfo{title}{{Congestion avoidance and control}}.
\newblock In \emph{\bibinfo{booktitle}{{Proceedings of SIGCOMM {\rq}88}}},
  \bibinfo{pages}{314--329} (\bibinfo{publisher}{ACM}, \bibinfo{year}{1988}).

\bibitem{Carvalho14}
\bibinfo{author}{Carvalho, R.} \emph{et~al.}
\newblock \bibinfo{title}{{Resilience of Natural Gas Networks during Conflicts,
  Crises and Disruptions}}.
\newblock \emph{\bibinfo{journal}{PLoS ONE}} \textbf{\bibinfo{volume}{9}},
  \bibinfo{pages}{e90265} (\bibinfo{year}{2014}).
  
\bibitem{Carvalho09}
\bibinfo{author}{Carvalho, R.}, \emph{et~al.}
\newblock \bibinfo{title}{{Robustness of trans-European gas networks}}.
\newblock \emph{\bibinfo{journal}{Phys. Rev. E}} \textbf{\bibinfo{volume}{80}},
\bibinfo{pages}{0161069} (\bibinfo{year}{2009}).  
  
\bibitem{Giridhar06}
\bibinfo{author}{Giridhar, A.} \& \bibinfo{author}{Kumar, P.~R.}
\newblock \bibinfo{title}{{Scheduling Automated Traffic on a Network of
  Roads}}.
\newblock \emph{\bibinfo{journal}{IEEE Transactions on Vehicular Technology}}
  \textbf{\bibinfo{volume}{55}}, \bibinfo{pages}{1467--1474}
  (\bibinfo{year}{2006}).

\bibitem{Helbing16}
\bibinfo{author}{Tachet, R.} \emph{et~al.}
\newblock \bibinfo{title}{{Revisiting Street Intersections Using Slot-Based
  Systems}}.
\newblock \emph{\bibinfo{journal}{PLOS ONE}} \textbf{\bibinfo{volume}{11}},
  \bibinfo{pages}{1--9} (\bibinfo{year}{2016}).

\bibitem{Yu2011}
\bibinfo{author}{Liu, Y.-Y.}, \bibinfo{author}{Slotine, J.-J.} \&
  \bibinfo{author}{Barab{\'a}si, A.-L.}
\newblock \bibinfo{title}{{Controllability of complex networks}}.
\newblock \emph{\bibinfo{journal}{Nature}} \textbf{\bibinfo{volume}{473}},
  \bibinfo{pages}{167--173} (\bibinfo{year}{2011}).

\bibitem{Devine14}
\bibinfo{author}{Devine, M.~T.}, \bibinfo{author}{Gleeson, J.~P.},
  \bibinfo{author}{Kinsella, J.} \& \bibinfo{author}{Ramsey, D.~M.}
\newblock \bibinfo{title}{{A Rolling Optimisation Model of the UK Natural Gas
  Market}}.
\newblock \emph{\bibinfo{journal}{Networks and Spatial Economics}}
  \textbf{\bibinfo{volume}{14}}, \bibinfo{pages}{209--244}
  (\bibinfo{year}{2014}).

\bibitem{Nepusz12}
\bibinfo{author}{Nepusz, T.} \& \bibinfo{author}{Vicsek, T.}
\newblock \bibinfo{title}{{Controlling edge dynamics in complex networks}}.
\newblock \emph{\bibinfo{journal}{Nature Physics}}
  \textbf{\bibinfo{volume}{8}}, \bibinfo{pages}{568--573}
  (\bibinfo{year}{2012}).

\bibitem{Szolnoki12}
\bibinfo{author}{Szolnoki, A.}, \bibinfo{author}{Perc, M.} \&
  \bibinfo{author}{Szabo, G.}
\newblock \bibinfo{title}{{Accuracy in strategy imitations promotes the
  evolution of fairness in the spatial ultimatum game}}.
\newblock \emph{\bibinfo{journal}{Epl}} \textbf{\bibinfo{volume}{100}},
  \bibinfo{pages}{28005} (\bibinfo{year}{2012}).

\bibitem{Bertsekas92}
\bibinfo{author}{Bertsekas, D.~P.} \& \bibinfo{author}{Gallager, R.}
\newblock \emph{\bibinfo{title}{{Data Networks}}} (\bibinfo{publisher}{Prentice
  Hall}, \bibinfo{year}{1992}).

\bibitem{Pioro04}
\bibinfo{author}{Pioro, M.} \& \bibinfo{author}{Medhi, D.}
\newblock \emph{\bibinfo{title}{{Routing, Flow, and Capacity Design in
  Communication and Computer Networks}}} (\bibinfo{publisher}{Morgan Kaufmann},
  \bibinfo{year}{2004}).

\bibitem{Srikant04}
\bibinfo{author}{Srikant, R.}
\newblock \emph{\bibinfo{title}{{The Mathematics of Internet Congestion
  Control}}} (\bibinfo{publisher}{Birkh{\"a}user}, \bibinfo{address}{Boston},
  \bibinfo{year}{2004}).

\bibitem{Albert02}
\bibinfo{author}{Albert, R.} \& \bibinfo{author}{Barabasi, A.~L.}
\newblock \bibinfo{title}{{Statistical mechanics of complex networks}}.
\newblock \emph{\bibinfo{journal}{Rev. Mod. Phys.}}
  \textbf{\bibinfo{volume}{74}}, \bibinfo{pages}{47--97}
  (\bibinfo{year}{2002}).

\bibitem{Boccaletti06}
\bibinfo{author}{Boccaletti, S.}, \bibinfo{author}{Latora, V.},
  \bibinfo{author}{Moreno, Y.}, \bibinfo{author}{Chavez, M.} \&
  \bibinfo{author}{Hwang, D.~U.}
\newblock \bibinfo{title}{{Complex networks: Structure and dynamics}}.
\newblock \emph{\bibinfo{journal}{Phys. Rep.-Rev. Sec. Phys. Lett.}}
  \textbf{\bibinfo{volume}{424}}, \bibinfo{pages}{175--308}
  (\bibinfo{year}{2006}).

\bibitem{CaldarelliBook07}
\bibinfo{author}{Caldarelli, G.}
\newblock \emph{\bibinfo{title}{{Scale-Free Networks: Complex Webs in Nature
  and Technology}}} (\bibinfo{publisher}{Oxford University Press},
  \bibinfo{address}{New York}, \bibinfo{year}{2007}).

\bibitem{HavlinBook10}
\bibinfo{author}{Cohen, R.} \& \bibinfo{author}{Havlin, S.}
\newblock \emph{\bibinfo{title}{{Complex Networks: Structure, Robustness and
  Function}}} (\bibinfo{publisher}{Cambridge University Press},
  \bibinfo{address}{New York}, \bibinfo{year}{2010}).

\bibitem{NewmanBook10}
\bibinfo{author}{Newman, M.}
\newblock \emph{\bibinfo{title}{{Networks: An Introduction}}}
  (\bibinfo{publisher}{Oxford University Press}, \bibinfo{address}{New York},
  \bibinfo{year}{2010}).

\bibitem{Kobayashi11}
\bibinfo{author}{Kobayashi, H.}, \bibinfo{author}{Mark, B.~L.} \&
  \bibinfo{author}{Turin, W.}
\newblock \emph{\bibinfo{title}{{Probability, Random Processes, and Statistical
  Analysis: Applications to Communications, Signal Processing, Queueing Theory
  and Mathematical Finance}}} (\bibinfo{publisher}{Cambridge University Press},
  \bibinfo{year}{2011}).

\bibitem{Lai05}
\bibinfo{author}{Zhao, L.}, \bibinfo{author}{Lai, Y.~C.},
  \bibinfo{author}{Park, K.} \& \bibinfo{author}{Ye, N.}
\newblock \bibinfo{title}{{Onset of traffic congestion in complex networks}}.
\newblock \emph{\bibinfo{journal}{Phys. Rev. E}} \textbf{\bibinfo{volume}{71}},
  \bibinfo{pages}{026125} (\bibinfo{year}{2005}).

\bibitem{Guimera02}
\bibinfo{author}{Guimer{\`a}, R.}, \bibinfo{author}{D{\'i}az-Guilera, A.},
  \bibinfo{author}{Vega-Redondo, F.}, \bibinfo{author}{Cabrales, A.} \&
  \bibinfo{author}{Arenas, A.}
\newblock \bibinfo{title}{{Optimal network topologies for local search with
  congestion}}.
\newblock \emph{\bibinfo{journal}{Physical Review Letters}}
  \textbf{\bibinfo{volume}{89}}, \bibinfo{pages}{248701}
  (\bibinfo{year}{2002}).

\bibitem{Guimera02a}
\bibinfo{author}{Guimera, R.}, \bibinfo{author}{Arenas, A.},
  \bibinfo{author}{Diaz-Guilera, A.} \& \bibinfo{author}{Giralt, F.}
\newblock \bibinfo{title}{{Dynamical properties of model communication
  networks}}.
\newblock \emph{\bibinfo{journal}{Phys. Rev. E}} \textbf{\bibinfo{volume}{66}},
  \bibinfo{pages}{026704} (\bibinfo{year}{2002}).

\bibitem{Cholvi05}
\bibinfo{author}{Cholvi, V.}, \bibinfo{author}{Laderas, V.},
  \bibinfo{author}{Lopez, L.} \& \bibinfo{author}{Fernandez, A.}
\newblock \bibinfo{title}{{Self-adapting network topologies in congested
  scenarios}}.
\newblock \emph{\bibinfo{journal}{Phys. Rev. E}} \textbf{\bibinfo{volume}{71}},
  \bibinfo{pages}{035103(R)} (\bibinfo{year}{2005}).

\bibitem{Arenas06}
\bibinfo{author}{Duch, J.} \& \bibinfo{author}{Arenas, A.}
\newblock \bibinfo{title}{{Scaling of fluctuations in traffic on complex
  networks}}.
\newblock \emph{\bibinfo{journal}{Physical Review Letters}}
  \textbf{\bibinfo{volume}{96}}, \bibinfo{pages}{218702}
  (\bibinfo{year}{2006}).

\bibitem{Stanley07}
\bibinfo{author}{Sreenivasan, S.}, \bibinfo{author}{Cohen, R.},
  \bibinfo{author}{Lopez, E.}, \bibinfo{author}{Toroczkai, Z.} \&
  \bibinfo{author}{Stanley, H.~E.}
\newblock \bibinfo{title}{{Structural bottlenecks for communication in
  networks}}.
\newblock \emph{\bibinfo{journal}{Phys. Rev. E}} \textbf{\bibinfo{volume}{75}},
  \bibinfo{pages}{036105} (\bibinfo{year}{2007}).

\bibitem{Danon08}
\bibinfo{author}{Danon, L.}, \bibinfo{author}{Arenas, A.} \&
  \bibinfo{author}{Diaz-Guilera, A.}
\newblock \bibinfo{title}{{Impact of community structure on information
  transfer}}.
\newblock \emph{\bibinfo{journal}{Phys. Rev. E}} \textbf{\bibinfo{volume}{77}},
  \bibinfo{pages}{036103} (\bibinfo{year}{2008}).

\bibitem{GuanrongChen09}
\bibinfo{author}{Yang, R.}, \bibinfo{author}{Wang, W.~X.},
  \bibinfo{author}{Lai, Y.~C.} \& \bibinfo{author}{Chen, G.~R.}
\newblock \bibinfo{title}{{Optimal weighting scheme for suppressing cascades
  and traffic congestion in complex networks}}.
\newblock \emph{\bibinfo{journal}{Phys. Rev. E}} \textbf{\bibinfo{volume}{79}},
  \bibinfo{pages}{026112} (\bibinfo{year}{2009}).

\bibitem{Chen15}
\bibinfo{author}{Chen, Y.~Z.} \emph{et~al.}
\newblock \bibinfo{title}{{Extreme events in multilayer, interdependent complex
  networks and control}}.
\newblock \emph{\bibinfo{journal}{Scientific Reports}}
  \textbf{\bibinfo{volume}{5}}, \bibinfo{pages}{17277--17277}
  (\bibinfo{year}{2015}).

\bibitem{Mo00}
\bibinfo{author}{Mo, J.~H.} \& \bibinfo{author}{Walrand, J.}
\newblock \bibinfo{title}{{Fair end-to-end window-based congestion control}}.
\newblock \emph{\bibinfo{journal}{IEEE-ACM Trans. Netw.}}
  \textbf{\bibinfo{volume}{8}}, \bibinfo{pages}{556--567}
  (\bibinfo{year}{2000}).

\bibitem{Ahuja93}
\bibinfo{author}{Ahuja, R.~K.}, \bibinfo{author}{Magnanti, T.~L.} \&
  \bibinfo{author}{Orlin, J.~B.}
\newblock \emph{\bibinfo{title}{{Network Flows: Theory, Algorithms, and
  Applications}}} (\bibinfo{publisher}{Prentice Hall}, \bibinfo{year}{1993}).

\bibitem{Boyd04}
\bibinfo{author}{Boyd, S.} \& \bibinfo{author}{Vandenberghe, L.}
\newblock \emph{\bibinfo{title}{{Convex Optimization}}}
  (\bibinfo{publisher}{Cambridge University Press}, \bibinfo{address}{New
  York}, \bibinfo{year}{2004}).

\bibitem{Kelly98}
\bibinfo{author}{Kelly, F.~P.}, \bibinfo{author}{Maulloo, A.~K.} \&
  \bibinfo{author}{Tan, D. K.~H.}
\newblock \bibinfo{title}{{Rate control for communication networks: shadow
  prices, proportional fairness and stability}}.
\newblock \emph{\bibinfo{journal}{Journal of the Operational Research Society}}
  \textbf{\bibinfo{volume}{49}}, \bibinfo{pages}{237--252}
  (\bibinfo{year}{1998}).

\bibitem{Carvalho12}
\bibinfo{author}{Carvalho, R.}, \bibinfo{author}{Buzna, L.},
  \bibinfo{author}{Just, W.}, \bibinfo{author}{Helbing, D.} \&
  \bibinfo{author}{Arrowsmith, D.~K.}
\newblock \bibinfo{title}{{Fair sharing of resources in a supply network with
  constraints}}.
\newblock \emph{\bibinfo{journal}{Phys. Rev. E}} \textbf{\bibinfo{volume}{85}},
  \bibinfo{pages}{046101} (\bibinfo{year}{2012}).

\bibitem{Chiu_and_Jain89}
\bibinfo{author}{Chiu, D.~M.} \& \bibinfo{author}{Jain, R.}
\newblock \bibinfo{title}{{Analysis of the increase and decrease algorithms for
  congestion avoidance in computer-networks}}.
\newblock \emph{\bibinfo{journal}{Computer Networks and ISDN Systems}}
  \textbf{\bibinfo{volume}{17}}, \bibinfo{pages}{1--14} (\bibinfo{year}{1989}).

\bibitem{Johnson15}
\bibinfo{author}{Johnson, S.~D.} \& \bibinfo{author}{D{\rq}Souza, R.~M.}
\newblock \bibinfo{title}{{Inequality and Network Formation Games}}.
\newblock \emph{\bibinfo{journal}{Internet Mathematics}}
  \textbf{\bibinfo{volume}{11}}, \bibinfo{pages}{253--276}.

\bibitem{Doyle02}
\bibinfo{author}{Low, S.~H.}, \bibinfo{author}{Paganini, F.} \&
  \bibinfo{author}{Doyle, J.~C.}
\newblock \bibinfo{title}{{Internet congestion control}}.
\newblock \emph{\bibinfo{journal}{IEEE Control Systems Magazine}}
  \textbf{\bibinfo{volume}{22}}, \bibinfo{pages}{28--43}
  (\bibinfo{year}{2002}).

\bibitem{Massoulie02}
\bibinfo{author}{Massoulie, L.} \& \bibinfo{author}{Roberts, J.}
\newblock \bibinfo{title}{{Bandwidth sharing: Objectives and algorithms}}.
\newblock \emph{\bibinfo{journal}{IEEE-ACM Trans. Netw.}}
  \textbf{\bibinfo{volume}{10}}, \bibinfo{pages}{320--328}
  (\bibinfo{year}{2002}).

\bibitem{Bertsimas10}
\bibinfo{author}{Bertsimas, D.}, \bibinfo{author}{Farias, V.~F.} \&
  \bibinfo{author}{Trichakis, N.}
\newblock \bibinfo{title}{{The Price of Fairness}}.
\newblock \emph{\bibinfo{journal}{Oper. Res.}} \textbf{\bibinfo{volume}{59}},
  \bibinfo{pages}{17--31} (\bibinfo{year}{2011}).

\bibitem{Carmi2007}
\bibinfo{author}{Carmi, S.}, \bibinfo{author}{Wu, Z.},
  \bibinfo{author}{L{\'o}pez, E.}, \bibinfo{author}{Havlin, S.} \&
  \bibinfo{author}{{Eugene Stanley}, H.}
\newblock \bibinfo{title}{{Transport between multiple users in complex
  networks}}.
\newblock \emph{\bibinfo{journal}{The European Physical Journal B}}
  \textbf{\bibinfo{volume}{57}}, \bibinfo{pages}{165--174}
  (\bibinfo{year}{2007}).

\bibitem{Carmi2008}
\bibinfo{author}{Carmi, S.}, \bibinfo{author}{Wu, Z.}, \bibinfo{author}{Havlin,
  S.} \& \bibinfo{author}{Stanley, H.~E.}
\newblock \bibinfo{title}{{Transport in networks with multiple sources and
  sinks}}.
\newblock \emph{\bibinfo{journal}{Epl}} \textbf{\bibinfo{volume}{84}},
  \bibinfo{pages}{28005} (\bibinfo{year}{2008}).

\bibitem{Tan99}
\bibinfo{author}{Tan, D. K.~H.}
\newblock \emph{\bibinfo{title}{{Mathematical Models of Rate Control for
  Communication Networks}}}.
\newblock Ph.D. thesis, \bibinfo{school}{Statistical Laboratory, University of
  Cambridge} (\bibinfo{year}{1999}).

\bibitem{Courant89}
\bibinfo{author}{Courant, R.} \& \bibinfo{author}{Hilbert, D.}
\newblock \emph{\bibinfo{title}{{Methods of Mathematical Physics}}},
  vol.~\bibinfo{volume}{1} (\bibinfo{publisher}{Wiley-Interscience},
  \bibinfo{year}{1989}).

\bibitem{Ball_PCM08}
\bibinfo{author}{Ball, K.}
\newblock \emph{\bibinfo{title}{{Optimization and Lagrange Multipliers}}},
  chap. \bibinfo{chapter}{III.64}, \bibinfo{pages}{255--257}
  (\bibinfo{publisher}{Princeton University Press}, \bibinfo{address}{New
  Jersey}, \bibinfo{year}{2008}).

\bibitem{Goh01}
\bibinfo{author}{Goh, K.~I.}, \bibinfo{author}{Kahng, B.} \&
  \bibinfo{author}{Kim, D.}
\newblock \bibinfo{title}{{Universal behavior of load distribution in
  scale-free networks}}.
\newblock \emph{\bibinfo{journal}{Physical Review Letters}}
  \textbf{\bibinfo{volume}{87}}, \bibinfo{pages}{278701}
  (\bibinfo{year}{2001}).

\bibitem{NacePioro08}
\bibinfo{author}{Nace, D.} \& \bibinfo{author}{Pioro, M.}
\newblock \bibinfo{title}{{Max-Min Fairness and Its Applications to Routing and
  Load-Balancing in Communication Networks: A Tutorial}}.
\newblock \emph{\bibinfo{journal}{IEEE Commun. Surv. Tutor.}}
  \textbf{\bibinfo{volume}{10}}, \bibinfo{pages}{5--17} (\bibinfo{year}{2008}).

\bibitem{Danila2006}
\bibinfo{author}{Danila, B.}, \bibinfo{author}{Yu, Y.}, \bibinfo{author}{Marsh,
  J.~A.} \& \bibinfo{author}{Bassler, K.~E.}
\newblock \bibinfo{title}{{Optimal transport on complex networks}}.
\newblock \emph{\bibinfo{journal}{Phys. Rev. E}} \textbf{\bibinfo{volume}{74}},
  \bibinfo{pages}{046106} (\bibinfo{year}{2006}).

\bibitem{Barthelemy11}
\bibinfo{author}{Barthelemy, M.}
\newblock \bibinfo{title}{{Spatial networks}}.
\newblock \emph{\bibinfo{journal}{Phys. Rep.-Rev. Sec. Phys. Lett.}}
  \textbf{\bibinfo{volume}{499}}, \bibinfo{pages}{1--101}
  (\bibinfo{year}{2011}).

\bibitem{Goh05}
\bibinfo{author}{Goh, K.~I.}, \bibinfo{author}{Noh, J.~D.},
  \bibinfo{author}{Kahng, B.} \& \bibinfo{author}{Kim, D.}
\newblock \bibinfo{title}{{Load distribution in weighted complex networks}}.
\newblock \emph{\bibinfo{journal}{Phys. Rev. E}} \textbf{\bibinfo{volume}{72}},
  \bibinfo{pages}{4} (\bibinfo{year}{2005}).

\bibitem{Piororeview14}
\bibinfo{author}{Ogryczak, W.}, \bibinfo{author}{Luss, H.},
  \bibinfo{author}{Pioro, M.}, \bibinfo{author}{Nace, D.} \&
  \bibinfo{author}{Tomaszewski, A.}
\newblock \bibinfo{title}{{Fair Optimization and Networks: A Survey}}.
\newblock \emph{\bibinfo{journal}{J. Appl. Math.}} \bibinfo{pages}{25}
  (\bibinfo{year}{2014}).

\bibitem{Wang08}
\bibinfo{author}{Wang, H.~J.}, \bibinfo{author}{Hernandez, J.~M.} \&
  \bibinfo{author}{{Van Mieghem}, P.}
\newblock \bibinfo{title}{{Betweenness centrality in a weighted network}}.
\newblock \emph{\bibinfo{journal}{Phys. Rev. E}} \textbf{\bibinfo{volume}{77}},
  \bibinfo{pages}{046105} (\bibinfo{year}{2008}).

\bibitem{Ullah98}
\bibinfo{author}{Ullah, A.} \& \bibinfo{author}{Giles, D. E.~A.}
\newblock \emph{\bibinfo{title}{{Handbook of Applied Economic Statistics}}}
  (\bibinfo{publisher}{CRC Press}, \bibinfo{address}{New York},
  \bibinfo{year}{1998}).

\end{thebibliography}
\end{document}